\documentclass[aps,a4paper,12pt]{revtex4}
\pdfoutput=1
\usepackage{float}
\usepackage{fancyhdr}
\usepackage{color}
\usepackage{graphicx}
\usepackage{epsfig} 
\usepackage{amssymb}
\usepackage{amsmath,amsfonts}
\usepackage{booktabs}
\usepackage{hyperref}
\usepackage{tikz}
\usetikzlibrary{shapes}
\usepackage{pgfplotstable}
\usepackage[footnotesize,singlelinecheck=off,justification=raggedright]{caption}
\usepackage{multirow}
\def\bra#1{\mathinner{\langle{#1}|}}
\def\ket#1{\mathinner{|{#1}\rangle}}

\def\prod#1#2{\mathinner{\langle{#1}|{#2}\rangle}}

\DeclareMathAlphabet{\mathbbmsl}{U}{bbm}{m}{sl}
{\catcode`\|=\active\gdef\Braket#1{\left<\mathcode`\|"8000\let|\bravert {#1}\right>}}
\def\bravert{\egroup\,\vrule\,\bgroup}
\def\Tr{\mathop{\mbox{\normalfont Tr}}\nolimits}

\def\CH{{\mathcal H}}

\def\NR{{\mathbb R}}
\def\NZ{{\mathbb Z}}

\def\pref#1{(\ref{#1})}

\DeclareRobustCommand\mysquare{\raisebox{0pt}{\tikz{\node[draw,scale=1,rectangle,fill=white](){};}}}
\DeclareRobustCommand\mycirclefilled{\raisebox{0pt}{\tikz{\node[draw,scale=0.65,circle, fill=black](){};}}}

\begin{document}

\title{
  On the complex behaviour of the density
  in composite quantum systems.
}

 \author{Filiberto Ares}
\email{fares@iip.ufrn.br}
\affiliation{International Institute of Physics, UFRN, 59078-970, Natal, RN, Brazil}
\author{Jos\'e G. Esteve}
 \email{esteve@unizar.es}
 \affiliation{Departamento de F\'{\i}sica Te\'orica, Universidad de Zaragoza,
50009 Zaragoza, Spain}
\affiliation{Instituto de Biocomputaci\'on y F\'{\i}sica de Sistemas
Complejos (BIFI)}   
\affiliation{Centro de Astropart\'\i culas y F\'\i sica de Altas Energ\'\i as (CAPA)
50009 Zaragoza, Spain}
  \author{Fernando Falceto}
\email{falceto@unizar.es}
 \affiliation{Departamento de F\'{\i}sica Te\'orica, Universidad de Zaragoza,
50009 Zaragoza, Spain}
\affiliation{Instituto de Biocomputaci\'on y F\'{\i}sica de Sistemas
Complejos (BIFI)}   
\affiliation{Centro de Astropart\'\i culas y F\'\i sica de Altas Energ\'\i as (CAPA)
50009 Zaragoza, Spain}
\author{Alberto Us\'on}
  \email{auson@ific.uv.es}
  \affiliation{Instituto de F\'{\i}sica Corpuscular (IFIC), CSIC 
  \& Universitat de Val\`encia, 46980 Valencia, Spain. }

%\date{\today; \quad Filename:\ {\tt\jobname.tex}}

%\begin{quote}
%{\tt  $\,$  \vskip -1.9cm         [Filename: \jobname.tex]}
%\end{quote}

\begin{abstract} 
In this paper, we study how the probability of presence
of a particle is distributed between the two parts of a
composite fermionic system. We uncover that the difference of
probability depends on the energy in a striking way and
show the pattern of this distribution. We discuss the main
features of the latter and explain analytically those that we
understand. In particular, we prove  that it is a
non-perturbative property and we find out a large/small coupling
constant duality. We also find and study features that may connect our problem with
certain aspects of non linear classical dynamics, like the existence of
resonances and sensitive dependence on the state of the system.
We show that the latter has indeed a similar origin than in classical
mechanics: the appearance of small denominators in the perturbative series.
Inspired by the proof of KAM theorem, we are able to deal with this problem
by introducing a cut-off in energies that eliminates these small denominators.
We also formulate some conjectures that we are
not able to prove at present but can be supported by numerical experiments.
  \end{abstract}

\maketitle

\section{Introduction}\label{sec_intro}

In spite of its apparent simplicity, many-body
quantum systems in one dimension have turned out
to be very useful models in order to understand 
and unravel many different phenomena, from entanglement
\cite{Latorre, Jin, Ares1, Ares2, Ares3, Ares8}
and quantum information \cite{Amico, Laflorencie} 
to new phases of matter \cite{Asboth, Bernevig, Viyuela, Viyuela2} and 
quantum chaos \cite{Alessio}. Moreover, the development 
of experimental techniques in cold atoms, ion traps and polarized 
molecules  has recently allowed to simulate these systems in the 
laboratory \cite{Bloch, Porras, Blatt, Atala, Micheli, 
Yan, Kuznetsova}.

Two of the most studied unidimensional many-body quantum systems
are the tight binding model and the Su-Schrieffer-Heeger
(SSH) model. The tight binding model consists of a lattice
with a fixed number of free fermions which can hop from
one site to the next one with a given probability.
In the simplest version of the  model, the sites of the lattice
(the position of the atoms)
are fixed and the hopping probability (the hopping integral)
is constant along the chain.
Physically, this system can be seen as a toy model for
a one-dimensional metal. It can also be mapped into the
XX spin chain via the Jordan-Wigner transformation.
When the atom vibrations are taken into account, the hopping probability
depends on the position of the nearest sites and, due to the Peierls theorem,
the chain dimerizes. In the Born-Oppenheimer approximation the
 hopping probabilities between the even-odd
and odd-even sites are different. This is the SSH model, which describes
a unidimensional insulator. It was firstly introduced to characterize
solitons in the polyacetylene molecule \cite{ssh, ssh2, Casahorran}. In last years,
the SSH model has attracted much attention since it displays
the essential properties of topological insulators \cite{Asboth, Bernevig}.

In this paper, we take the union of two different systems 
of this type. That is, we analyze systems composed by two different
tight binding models coupled by special bonds which we will call
contacts. Physically, this situation corresponds to the junction
of two metals with different band structure.
We may also combine a tight binding model and a SSH model (metal-insulator)
or two SSH models (insulator-insulator). These kinds of junctions
were considered in Ref. \cite{Eisler}, in which the ground state
entanglement entropy between the two parts is investigated;
see also \cite{Peschel, Sakai}. 
Systems with two different critical parts (such as the tight 
binding model) or with a critical and a non-critical part 
(like the SSH model) have been examined from the perspective
of conformal invariance \cite{Hinrichsen, Berche, Zhang1, Zhang2}. 
Composite free-fermionic systems are also of interest in quantum 
transport and non equilibrium physics \cite{Antal, Barnabe, Biella, Calabrese, 
Eisler2, Stephan, Kennes, Viti, Allegra, Dubail, Biella2, Prosen, Gawedzki, 
Gawedzki2, Moosavi}, where a typical problem is 
the analysis of the evolution of the overall state 
of two different chains after being joined together (inhomogeneous 
quench).

Here, we consider the one particle states with a definite energy.
Depending on its energy, the particle is confined in one of the two
parts or, on the contrary, is delocalized along the whole chain.
In this work, we will analyze how the particle distributes between
the two parts. In a way, our problem is not {\bf how}, but {\bf where} 
Schr\"odinger's cat is. For this purpose, we will introduce a quantity that
we call \textit{leaning}, defined as the difference between the 
probabilities of finding the particle in each part of the chain.
It happens that the dependence of the leaning on the energy and 
on the contact between the two parts is rather non-trivial \cite{Uson}. The 
goal of this paper is to characterize and explain this behaviour.

The paper is organized as follows. In Section \ref{sec_basic}, we introduce
the main system under study, the union of two tight binding models,
and the so-called leaning. In Section \ref{sec_analytic}, we will see how
to compute analytically the leaning of a one-particle configuration.
In Section \ref{sec_spectrum},
we calculate the spectral density of the whole chain. Sections \ref{sec_resonance} and
\ref{sec_boundary} are devoted respectively to analyze the resonant regions and 
to determine the boundary of the clouds of points that appear in 
the energy-leaning plot. In Section \ref{sec_measure}, we conjecture
the existence of a measure in the energy-leaning plane that accounts
for the density of points in the thermodynamic limit. Finally,
in Section \ref{sec_conclu}, we present our conclusions and outlook.
The paper is complemented with an Appendix where we show that the average
of the leaning does not depend on the value of the contact. 

\section{Basic set-up}\label{sec_basic}

As we already mentioned in the introduction, the system that 
we are going to study consists of the union of two
tight binding fermionic chains of lengths $N_1$ and $N_2$ 
with hopping parameters $t_1$ and $t_2$ respectively.
The ends of the two chains are connected by means of other
hoppings $t_0,t_0'$ which we call contacts, these
will be our main tunable parameters.

Therefore, the Hamiltonian of the composite system is
\begin{eqnarray}\label{hamil}
H=\frac12
\left(
t_1
\sum_{n=1}^{N_1-1}
(
a_n^\dagger a_{n+1}
+
a_{n+1}^\dagger a_{n}
)
+
t_2
\sum_{m=1}^{N_2-1}
(
b_m^\dagger b_{m+1}+
b_{m+1}^\dagger b_{m}
)\hskip 0.5cm
\right.
\cr
\left.
+
t_0
(
a_{N_1}^\dagger b_{N_2}
+
b_{N_2}^\dagger a_{N_1}
)
+
t'_0
(
b_1^\dagger a_1
+
a_1^\dagger b_1
)
\right),
\end{eqnarray}
where $a_n, b_m$ are the fermionic annihilation operators associated
to every piece of the composite chain. Note that the sites 
are enumerated such that the site $n=1$ of the subchain with 
hopping $t_1$ is connected with the site $m=1$ of the subchain 
with hopping $t_2$ by the contact coupling $t_0'$ and, likewise, the site 
$n=N_1$ is connected with the site $m=N_2$ by the contact $t_0$
(that is, the sites of the subchain with hopping $t_2$
are numbered in opposite direction to those of the subchain
with hopping $t_1$).

This kind of systems has some interesting properties
that deserve further research.
For instance, at certain values of the contacts it possesses
a discrete spectrum with localized, topologically protected states.
They behave similarly to those of the topological insulators.
These features will be studied elsewhere. Here we are
rather interested in the continuous spectrum
(in the thermodynamic limit) and more precisely in
its one particle states
\begin{equation}\label{one_particle}
\Psi=\left(\sum_{n=1}^{N_1}  \alpha_n a^\dagger_n+
\sum_{m=1}^{N_2} \beta_m b^\dagger_m\right)\ket0,
\end{equation}
where $\ket{0}$ represents the vacuum of the Fock
space, i.e. $a_n\ket0=b_m\ket0=0$ $\forall m, n$.

\begin{figure}
 \center
  \includegraphics[width=10cm]{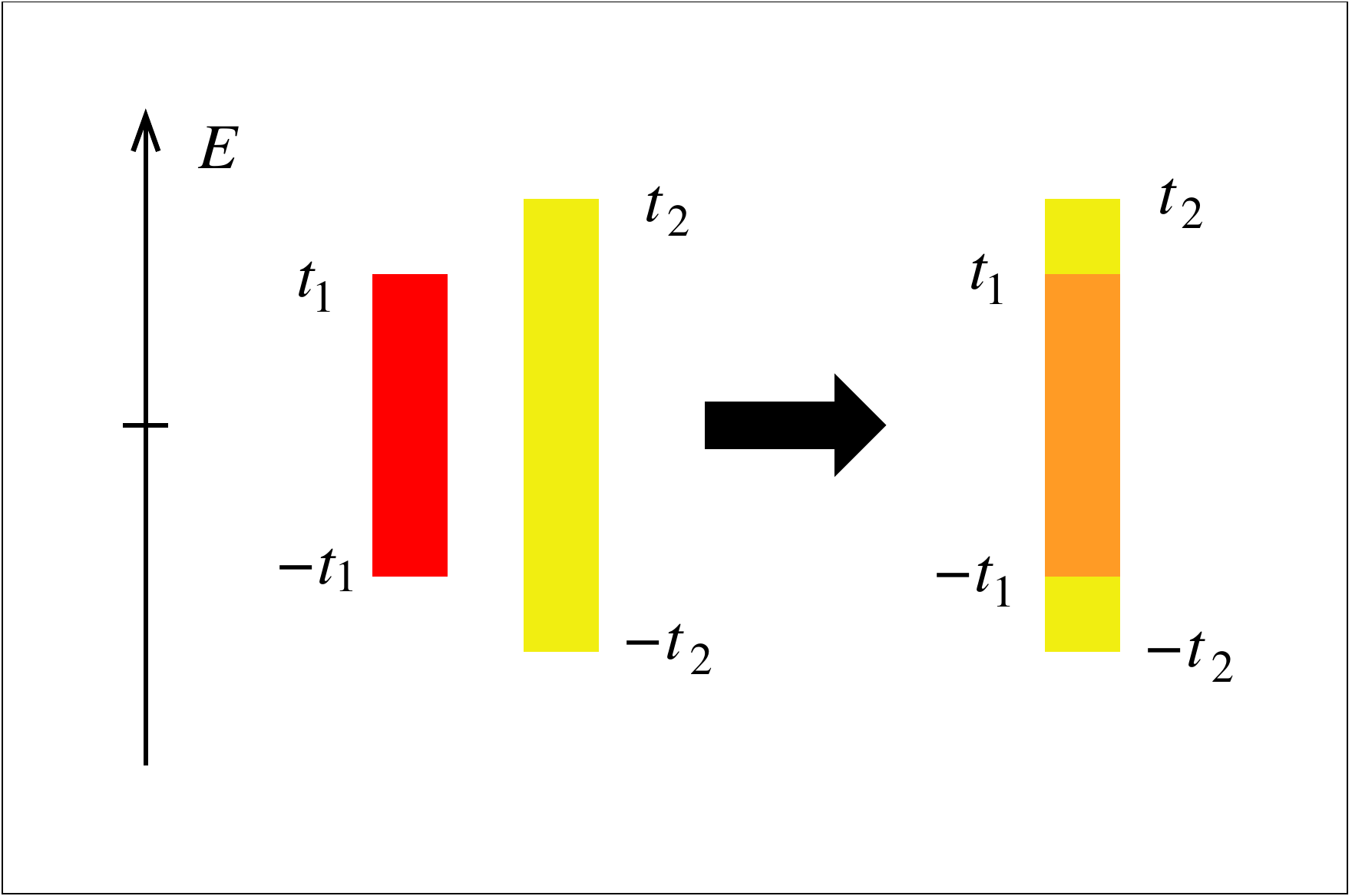}
  \caption{In the figure we represent the band structure of one particle
    states for the two separate subsystems (on the left) and
    the combined spectrum when we connect them (on the right).
    In the latter case and for particular values of the contacts
    $t_0, t_0'$ there could be localized states with energy in the discrete
    spectrum,  outside the band.
    Here we are not interested in these states and we do not represent them.} 
\label{bands}
 \end{figure}

The continuous spectrum in the composite system has
a band structure that is obtained as a superposition
of those corresponding to every of its two parts.
In fig. \ref{bands} we represent this situation.

For definiteness and without lose of generality, we shall take
$t_2>t_1>0$. Then, the  states whose energies are such
$t_2>|E|>t_1>0$ are mainly supported in the region with hopping parameter
$t_2$ and hardly penetrate, with exponential decay, in the left
hand side. On the contrary, those states with energy in the interval
$[-t_1,t_1]$ are distributed along the whole chain. Our concern in this work
is how the latter split between the two parts of the chain. 

Therefore, we decompose the one particle Hilbert space
$$\CH=\CH_1\oplus\CH_2,$$
where $\CH_1$ contains the wave functions supported
in the left hand side ($\beta_m=0$) and 
$\CH_2$ those supported in the right hand side
($\alpha_n=0$) and denote by $P_1$ and $P_2$ the
corresponding orthogonal projectors;
we shall be interested in the expectation value of the difference
between these projectors
$${L}={\bra\Psi(P_2-P_1)\ket\Psi}.$$

We  will refer to ${L}$ as the {\it leaning}.
It measures the difference between the probability
of finding the particle in the right hand side
and that of finding it in the left one.

\begin{figure}
\hskip -.7cm\includegraphics[width=8.5cm]{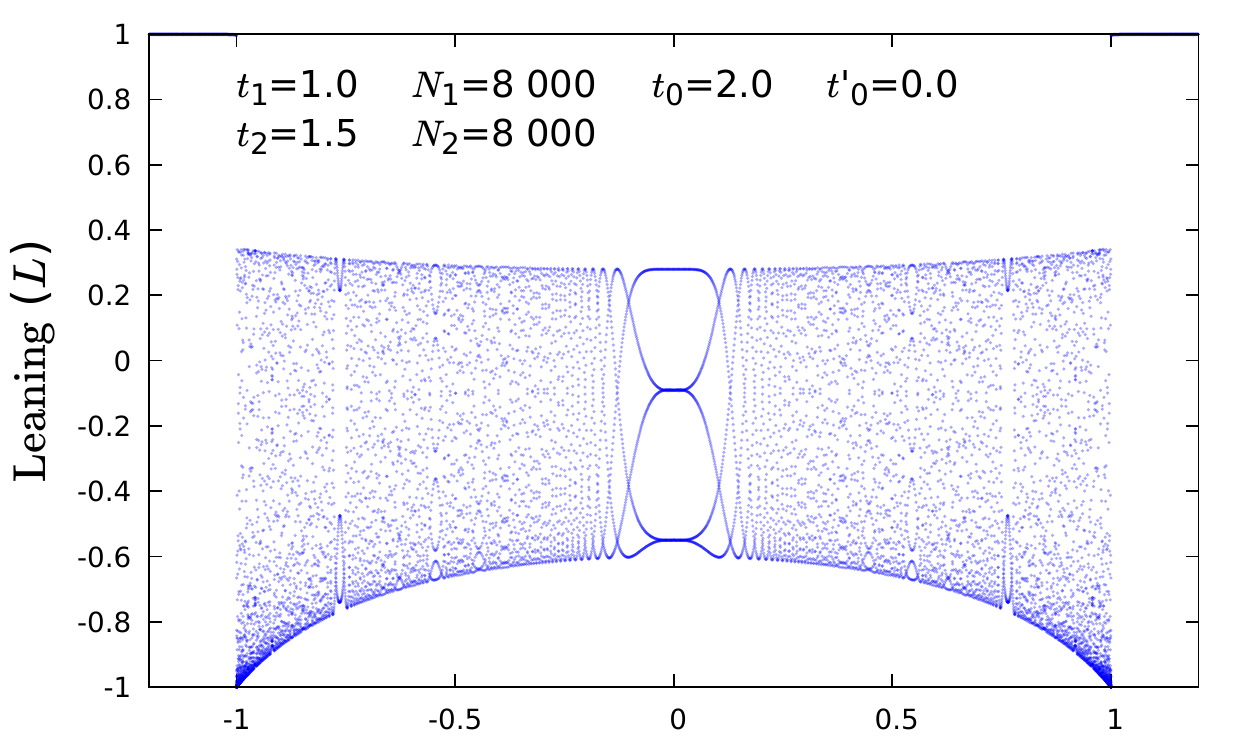}
\hskip -.7cm\includegraphics[width=8.5cm]{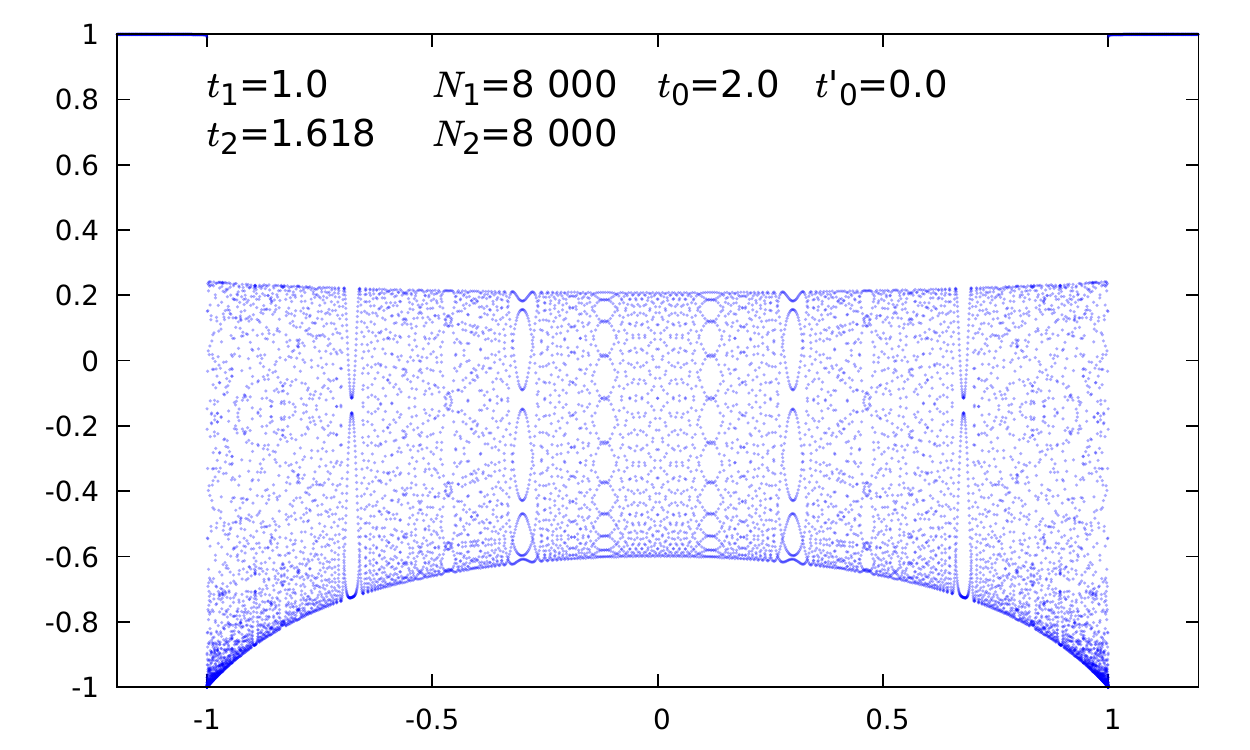}

\hskip -.7cm\includegraphics[width=8.5cm]{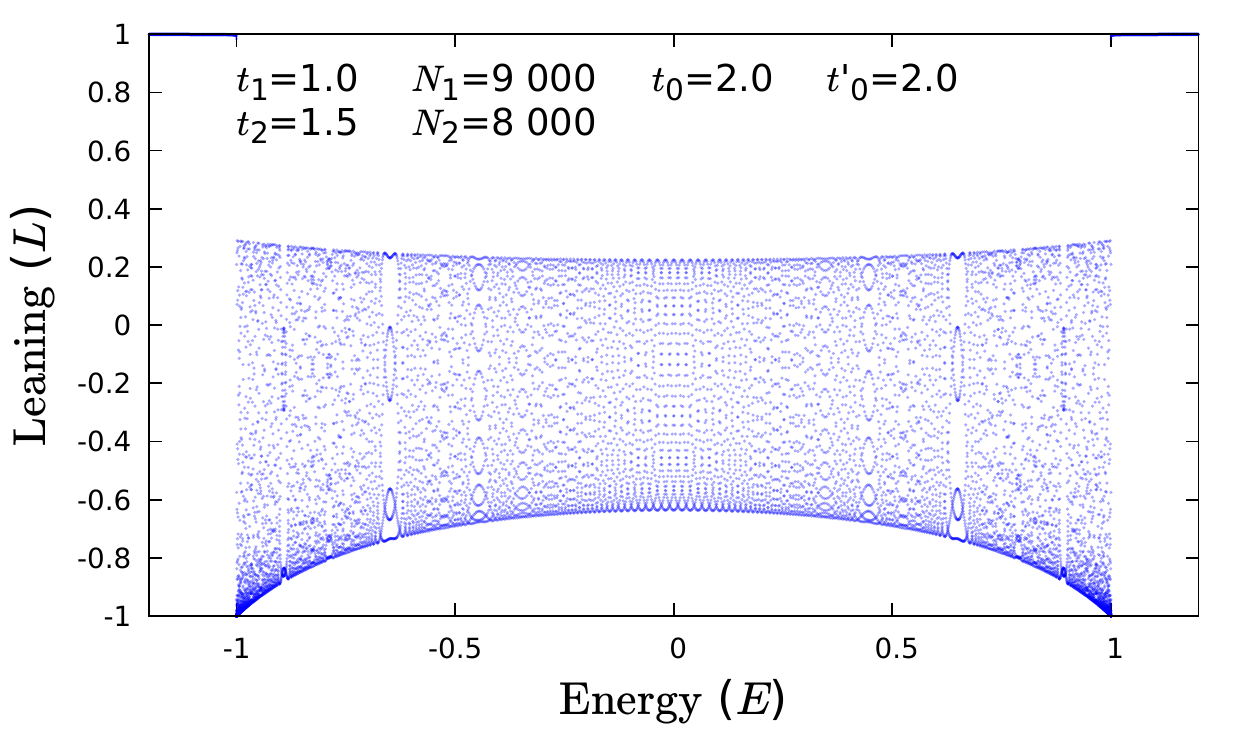}
\hskip -.7cm\includegraphics[width=8.5cm]{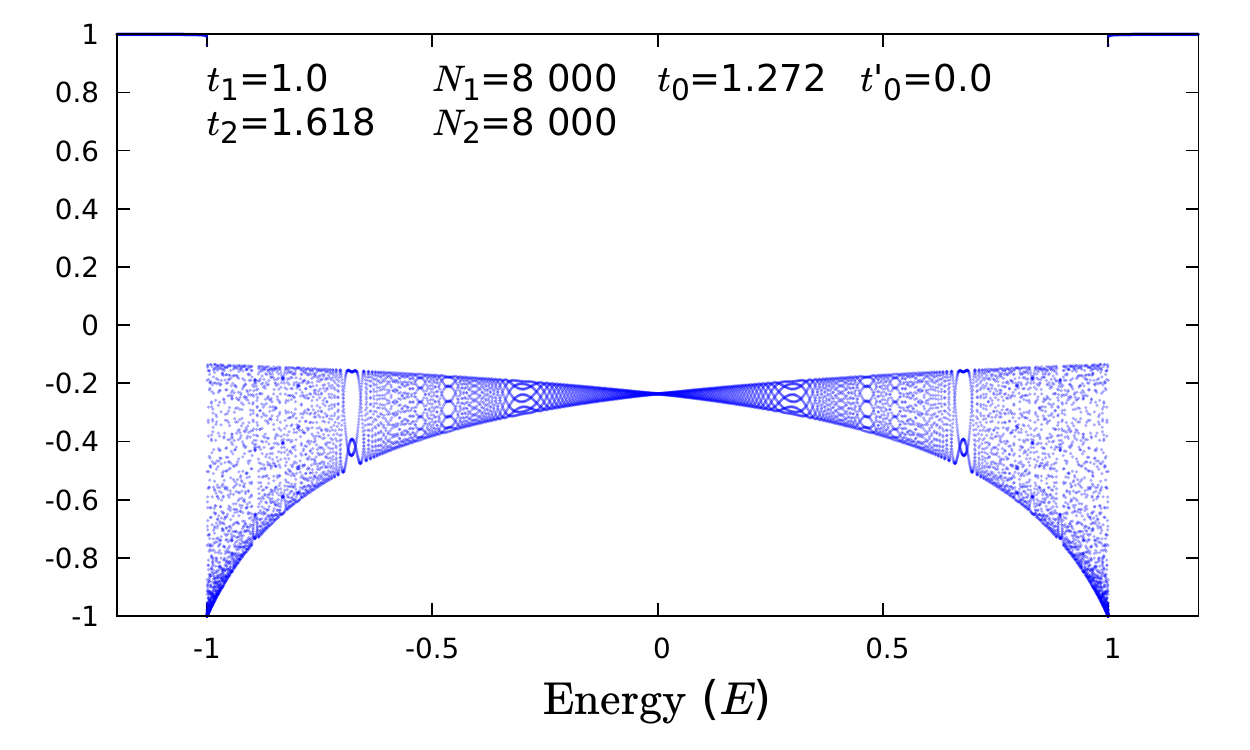}
\caption{
  In these plots we show the dependence of the leaning (in the vertical axis)
  as a function of the energy (in horizontal axis) for different values of the
  hoppings $t_1$ and  $t_2$, sizes $N_1$ and $N_2$ of the two subsystems, and 
  contacts $t_0$, $t_0'$.
  The two plots on the left differ in the size of one of the subsystems $N_1$.
  The two on the right in the value of the contact $t_0$.}
\label{intro}
\end{figure}

The leaning associated to a one particle stationary state
$\Psi_E$ will be denoted by ${L}_E$. 
It is immediate to see that, in the thermodynamic limit,
${L}_E=1$ for $|E|\in[t_1,t_2]$,
however, when $E\in[-t_1,t_1]$ the leaning depends on the energy
in a rather complex way, as it is shown in fig. \ref{intro}.

The rest of the paper is devoted to understand this behaviour
as fully as possible.

In first place, we see a cloud of points that fills a definite region
in the $E$-${L}$ plane, apparently bounded by smooth curves.
The cloud is symmetric under the exchange of $E$ to $-E$.
We also observe that the points seem to be randomly distributed 
inside the region, except for some range of energy where they 
group at some definite values of ${L}$ and lie along visible curves. 
We call these zones resonant. We will study the origin of 
the previous facts and how they depend on the parameters and 
size of the system. Further properties of these plots will be 
discussed along this work.

\section{Analytic approach.}\label{sec_analytic}

In this section we will show how to determine analytically the leaning of
a given eigenstate.

The coefficients $\alpha_n$ and $\beta_n$ for an
eigenfunction of the Hamiltonian
with energy $E$, such that $|E|<t_1<t_2$, can be written as
$$\alpha_n=A_1\sin(n{k}_1+\delta_1),\quad \beta_n=A_2\sin(n{k}_2+\delta_2),
$$
where
$${k}_j(E)=\arccos\frac{E}{t_j},\quad j=1,2,$$
and the following gluing conditions should be fulfilled
\begin{eqnarray}\label{gluing}
A_1t_1\sin(N_1{k}_1+{k}_1+\delta_1) -
A_2t_0\sin(N_2{k}_2+\delta_2)=0,
\cr
A_1t_0\sin(N_1{k}_1+\delta_1) -
A_2t_2\sin(N_2{k}_2+{k}_2+\delta_2)=0,\cr
A_1t_1\sin(\delta_1) -
A_2t'_0\sin({k}_2+\delta_2)=0,
\cr
A_1t'_0\sin({k}_1+\delta_1) -
A_2t_2\sin(\delta_2)=0.
\end{eqnarray}
From the compatibility of these equations we derive
the spectral condition and henceforth the allowed values
for $E$. 

In order to simplify the analysis, we shall take
$t'_0=0$ and, therefore, the last two equations
imply $\delta_1=\delta_2=0$.
We shall show along the paper that this assumption
does not affect, in fact, the generality of our results.

Hence the other two equations can be equivalently written
\begin{equation}\label{main_eq}
C\equiv A_2/A_1=\frac{t_1}{t_0}\frac{\sin(N_1{k}_1+{k}_1)}{\sin(N_2{k}_2)}
=\frac{t_0}{t_2}\frac{\sin(N_1{k}_1)}{\sin(N_2{k}_2+{k}_2)},
\end{equation}
and finally
\begin{equation}\label{leaning}
  {L}=\frac{N_1-N_2C^2}{N_1+N_2C^2}
\end{equation}
is determined once the value of $E$ is fixed.

Equation \pref{main_eq} is our starting point for the study
of plots in fig. \ref{intro}. In the next sections we will 
focus on different aspects or characteristics of these plots 
and we will show how they emerge from \pref{main_eq}. But 
before going to that, we shall insert a paragraph
to explain the symmetry under the exchange of $E$ with $-E$
that we observe in the plots.

This is due to a chiral transformation in the states that
reverses the sign of the Hamiltonian. Namely, for a one 
particle state (\ref{one_particle}),
we define its chiral transformed state by
$$\Gamma\Psi=
\left(\sum_{n=1}^{N_1}  (-1)^n
\alpha_n a^\dagger_n-
\sum_{m=1}^{N_2} (-1)^m\beta_m b^\dagger_m\right)\ket0.
$$
If $N_1+N_2$ is even or $t_0t_0'=0$, the Hamiltonian (\ref{hamil})
satisfies $\Gamma H\Gamma=-H$ while, for the projectors $P_1$, $P_2$,
we have $\Gamma P_i\Gamma=P_i$.
That is, the chiral symmetry reverses the sign of the energy
without changing the leaning of the state. This explains
the symmetry in the plots of fig. \ref{intro}.

\section{Spectral density}\label{sec_spectrum}

In this section we compute the spectral density
$\lambda_{t_0}(E)$ of the composite system in the thermodynamic limit
(when the size of the chain goes to infinity) while keeping the relative
size of the two chains fixed, $N_1=\nu_1 N$ and $N_2=\nu_2 N$. In order to make
the computation simpler, we will take $t_0'=0$. 
 
Call $\Sigma_{t_0,N}$ the spectrum of $H$ for given values of $t_0$ and $N$,
($t_1, t_2, \nu_1, \nu_2$ remain fixed and, to simplify the notation, are omitted
in the symbol used for the spectrum). We are interested in the
region of the spectrum where the bands of the two pieces of the chain overlap,
i. e.  $E\in(-t_1,t_1)$. We define the spectral density as
\begin{equation}\label{spec_density}
  \lambda_{t_0}(E)=\lim_{\delta E\to0^+}\frac1{2\,\delta E}
\lim_{N\to\infty}N^{-1}
\,\sharp\!\left(\Sigma_{t_0,N}\cap [E-\delta E, E+\delta E]\right),
\end{equation}
where the symbol $\sharp$ stands for the cardinality of the set.
% and $\Omega_N = \sharp(\Sigma_{t_0,N}\cap [-t_1, t_1])$
% is the appropriate normalization to make
% $\lambda_{t_0}$ a probability density
% in $(-t_1,t_1)$. 

We will show that the density of states
is actually independent of $t_0$ and can be computed
by simply adding up the density of the two pieces.
The latter can be easily estimated by going from $k$ space,
where the points in the spectrum are regularly spaced by intervals
$\pi/(N_i+1)$,
to the $E$ space. As a result one gets
%  $$\displaystyle \lambda_{t_0}(E)=\frac{\nu_2/{\sqrt{t_2^2-E^2}}+\nu_1/{\sqrt{t_1^2-E^2}}} {2\nu_2\arcsin(t_1/t_2)+\nu_1\pi}.$$
\begin{equation}\label{unorm_dens}  
  \displaystyle \lambda_{t_0}(E)=
  \frac{\nu_1k'_1(E)+\nu_2k'_2(E)}{\pi}=\frac{\nu_1/{\sqrt{t_1^2-E^2}}+
      \nu_2/{\sqrt{t_2^2-E^2}}} {\pi}.
\end{equation}    

One might be tempted to approach the problem
by using perturbation theory. Actually,
if we decompose the Hamiltonian in (\ref{hamil})
as $H=H_0+H_I$ with $H_0$  the unperturbed piece and the
perturbation given by the contact term
$$H_I=  
\frac{t_0}{2}
(
a_{N_1}^\dagger b_{N_2}
+
b_{N_2}^\dagger a_{N_1}
)
+
\frac{t'_0}{2}
(
b_1^\dagger a_1
+
a_1^\dagger b_1
),
$$
one immediately sees that for $\varphi_k, \varphi_{k'}$ eigenstates of $H_0$
$$\bra{\varphi_k}H_I\ket{\varphi_{k'}}=O\left(\frac1N\right),$$
and, therefore, the interaction between the two pieces decreases
when the system gets larger.
While this is true, if we try to apply the perturbative expansion
we have to face a sort of  ``small denominators'' problem, well known in
classical perturbation theory. In fact, when $N$ grows,
the gaps $E^0_m-E^0_{m'}$, which appear in the denominators of the
perturbative expansion,
can be arbitrarily small (even smaller than $O(1/N)$ for certain
values of $m,m'$) and the perturbative expansion ceases to make sense. 

Another indication that we are dealing with a non perturbative
phenomenon is the fact that, while for $t_0,t_0'=0$ 
the only two possible values
for the leaning are 1 or $-1$, for any value of $t_0\not=0$
we have (with $N$ large enough) states with an arbitrary value
for ${L}$ in the interval $[-1,1]$.
This means that we can not approximate perturbatively
the states of the composite system.

To avoid these potential problems we shall take a non perturbative
avenue to estimate the energy eigenvalues.

We are interested in the case $t'_0=0$ in which the equations (\ref{main_eq}) apply
and the spectrum $\Sigma_{t_0,N}$ is given by the solutions for $E$ of
the equation
\begin{equation}\label{spectral_eq}
  \frac{t_0^2}{t_1t_2}=
\frac{\sin(N_1{k}_1+{k}_1)\sin(N_2{k}_2+{k}_2)}
{\sin(N_1{k}_1)\sin(N_2{k}_2)}.
\end{equation}
Recall that $k_i$ and the energy are related by
$E=t_i\cos k_i$. 

In  the two extreme limits, $t_0=0$ and $t_0\to\infty$, it is easy to determine
$\Sigma_{t_0,N}$. Actually, for $t_0=0$,  $\Sigma_{0,N}$ is
the union of the spectra of the two parts of the composite system with
Dirichlet boundary conditions, that is
\begin{eqnarray*}
  \Sigma_{0,N}=\{E^0_m;\ m=1,\dots,N_1+N_2\}=&&\\[2mm]&&\hskip -6cm
  =\left\{t_1\cos\left(\frac{\pi m_1}{N_1+1}\right);\ m_1=1,\dots,N_1\right\}\cup
  \left\{t_2\cos\left(\frac{\pi m_2}{N_2+1}\right);\ m_2=1,\dots,N_2\right\}.
\end{eqnarray*}

Similarly for $t_0\to\infty$ the spectrum is given by
\begin{eqnarray*}
  \Sigma_{\infty,N}=\{E^\infty_m;\ m=1,\dots,N_1+N_2-2\}=&&\\[2mm]&&\hskip -7.5cm
  =\left\{t_1\cos\left(\frac{\pi m_1}{N_1}\right);\ m_1=1,\dots,N_1-1\right\}\cup
  \left\{t_2\cos\left(\frac{\pi m_2}{N_2}\right);\ m_2=1,\dots,N_2-1\right\}.
\end{eqnarray*}

One may notice that in the latter case the spectrum has two points less than
for $t_0=0$. Actually, the missing eigenvalues correspond,
for large but finite $t_0$, to states localized at the contact
whose energies, close to $\pm t_0$, lie outside the spectral band.
When $t_0$ goes to infinity the energy of these states diverges.

To understand the spectrum for intermediate values of $t_0$ and $E\in(-t_1,t_1)$
(recall that we assume $t_2>t_1>0$) it is convenient
to write (\ref{spectral_eq}) in the form
$$t_0^2=f_1(E)f_2(E),$$
where
$$f_i(E)=E-\sqrt{t_i^2-E^2}\cot(N_i k_i(E)), \quad i=1, 2.$$

The crucial observation now is that
$$f_i'(E)= 1+N_i-\frac E{\sqrt{t_i^2-E^2}}\cot(N_i k_i(E))+
N_i\cot^2(N_i k_i(E))$$
is positive provided
$$t_i^2-E^2>\frac{E^2}{4N_i(N_i+1)}.$$
This means that for any $E$ in the open interval
$(-t_1,t_1)$ and $N_i=\nu_iN$, as before, there exists a
$K$  such that $f_i'(E)>0$ for any $N>K$.
As we are interested in the large $N$ limit, we may assume
that  $f_i'(E)$ is positive for any $E\in(-t_1,t_1)$.

\begin{figure}
  \includegraphics[width=14cm]{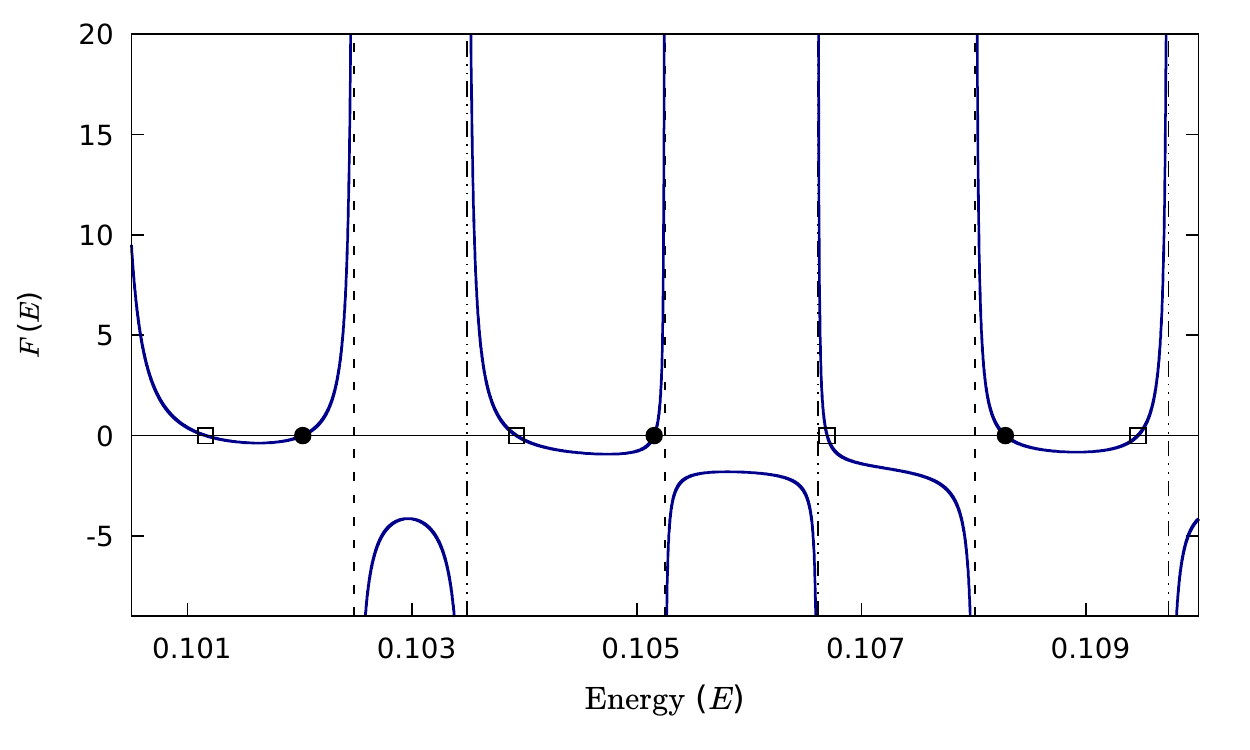}
  \caption{In this plot we represent the function $F(E)\equiv f_1(E)f_2(E)$ whose level set
    $F(E)=t_0^2$ is the spectrum. The parameters of the system are
    $t_1=1, t_2=1.5, N_1=1000, N_2=1700$. For clarity of the plot
    we have selected a short range of energies between 0.1 and 0.11. Notice
    that, as we prove in the
    text, the function is monotonic for intervals in which it is positive.
    We indicate by $\mycirclefilled$ and $\mysquare$  the zeros of $f_1(E)$
    and $f_2(E)$ respectively (which correspond to the energies $E_m^0$) 
    while the vertical lines are their asymptotes (associated to the energies
    $E_m^\infty$): the dot-dashed ones correspond to $f_1$ and the dashed to $f_2$.}
\label{spectrum}
\end{figure}

A consequence of the previous fact is that $F(E)\equiv f_1(E)f_2(E)$
is monotonic in every interval in which $F(E)>0$. Then the picture we get is
represented in fig. \ref{spectrum}.

For a point in $\Sigma_{0,N}$, say  $E^0_m$, we have $F(E^0_m)=0$.
Now assume $F'(E^0_m)>0$, then for $E$ slightly larger than $E^0_m$,
$F(E)>0$ and according to the previous result $F'(E)>0$. Therefore,
$F(E)$ increases with $E$ until we encounter an eigenvalue of $H$ for
$t_0=\infty$, say $E^\infty_{m'}$. At this point $F(E)$ diverges.
This implies that for each value of $t_0$ we have one and only one
solution for the equation (\ref{spectral_eq}) with energy in the interval
$[E^0_m,E^\infty_{m'}]$. In the case $F'(E^0_m)<0$ we have a similar result
but now going down in energies in such a way that there is one
and only one eigenstate of $H$ for each value of $t_0$ with
energy in the interval
$[E^\infty_{m'},E^0_m]$, where $E^\infty_{m'}$ is the point in $\Sigma_{\infty,N}$
immediately smaller than $E^0_m$.
Finally, in the unlikely instance in which
$F'(E^0_m)=0$ one necessarily has $f_1(E^0_m)=f_2(E^0_m)=0$ and therefore
$F''(E^0_m)=f'_1(E^0_m)f'_2(E^0_m)>0$. But this later property means that
$F(E)>0$ for $E$ in a punctured neighbourhood of $E^0_m$, hence the arguments
above hold and $F(E)$ grows monotonically to infinity when
we separate from $E^0_m$ in both directions
and approach the immediate points of $\Sigma_{\infty,N}$.

Due to this fact, we clearly see that given an
interval of energies $I\subset(-t_1,t_1)$
the number of stationary states with energies in $I$ varies at most by two
with $t_0$, namely
$$2\geq\sharp(I\cap\Sigma_{t_0})-\sharp(I\cap\Sigma_{0})\geq-2.$$
Therefore, the density of states $\lambda_{t_0}$
derived from (\ref{spec_density}) is independent of $t_0$
and, as we anticipated at the beginning
of this section, it can be written
as the sum of the densities for the two chains, i. e.
\begin{equation}\label{frequency}
\displaystyle \lambda_{t_0}(E)=
\frac{\nu_1/{\sqrt{t_1^2-E^2}}+\nu_2/{\sqrt{t_2^2-E^2}}}
{\pi}.
\end{equation}
  For further purposes we also introduce the spectral density normalized
  in the interval of energies $(-t_1,t_1)$ in which the stationary states
  extend along the whole chain
  $$\displaystyle \hat\lambda_{t_0}(E)=\frac{\nu_1/{\sqrt{t_1^2-E^2}}+\nu_2/{\sqrt{t_2^2-E^2}}}
  {\pi\nu_1+2\nu_2\arcsin(t_1/t_2)}.$$

\begin{figure}
 \center
  \includegraphics[width=14cm]{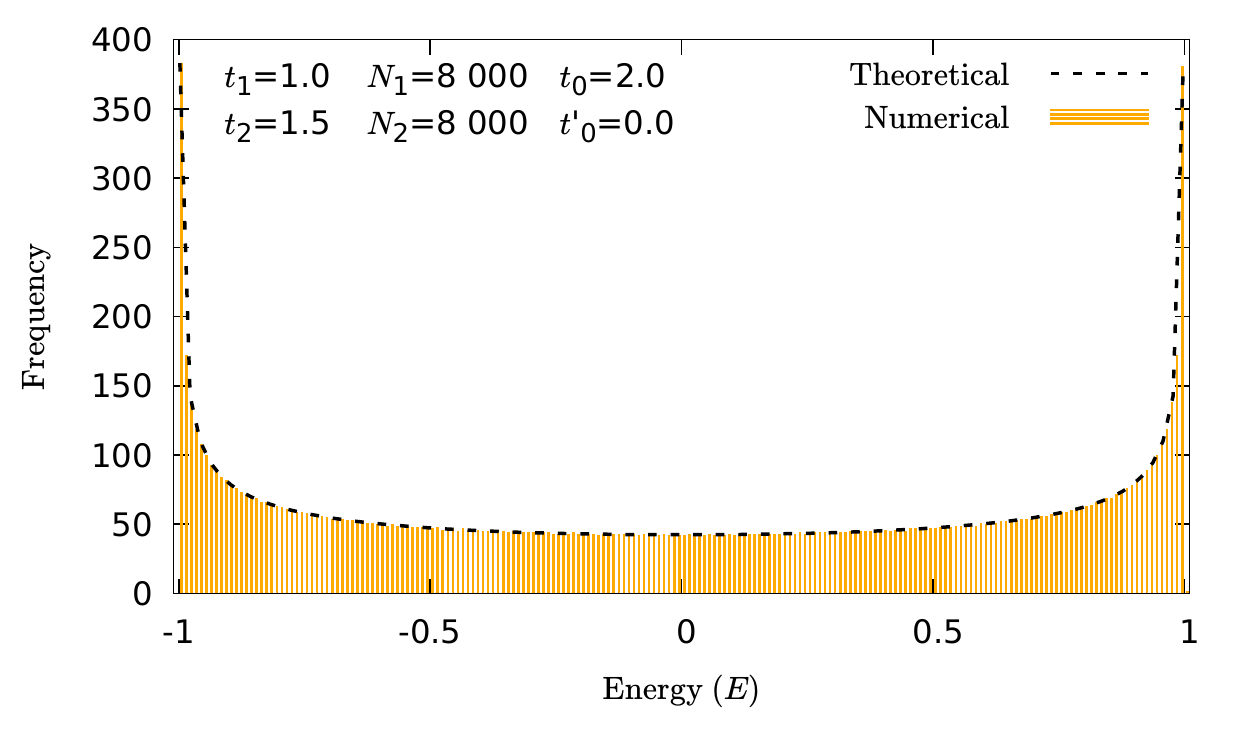}
  \caption{In the histogram we represent the frequency of the spectrum with
    bins of width $0.01$ units of energy.
    The discontinuous line is the theoretical
    prediction obtained from (\ref{frequency}).}
\label{histo_energy}
 \end{figure}

  This result has been checked numerically and the
  results are presented  in fig. \ref{histo_energy}.
  There the histogram for the energy of the states, determined numerically, is plotted
 against the theoretical curve obtained above.
 It is quite manifest the perfect agreement of both results.

Notice that we have shown, in passing, that for $E\in(-t_1,t_1)$,
$t_0\not=0$ or $\infty$ and
$N$ large enough, there is no degeneracy in the spectrum of $H$. This is
very good news because in the case of a degenerate eigenvalue the
leaning of the eigenstates is not well defined; it means that
the accidental degeneracy (and the undefinition of the leaning)
can only possibly happen in the extreme cases $t_0=0$ or $\infty$,
and this only after fine tunning $t_1,t_2$ and $N$.

\section{Resonant regions}\label{sec_resonance}

We turn now our attention to other feature of the plots
in fig. \ref{intro}: the existence of resonances.

We call resonant regions those windows in energy
where the points in the plot accumulate at a few
definite values of the leaning, like for instance in the central
zone of the upper-left panel of fig. \ref{intro}.
Outside these regions the points $(E,{L}_E)$ spread and the cloud seems
to fill the whole allowed band.
We have observed that the width of the resonances depends on $N$,
in such a way that they shrink when $N$ grows.

To understand the reason for these facts let us
consider a solution $E_0$ for \pref{main_eq} 
$$
C_0= \frac{t_1}{t_0}\frac{\sin((N_1+1){k}_1(E_0))
}{\sin(N_2{k}_2(E_0))}=\frac{t_0}{t_2}
\frac{\sin(N_1{k}_1(E_0))}{\sin((N_2+1){k}_2(E_0))},
$$
and expand around this value
\begin{eqnarray}\label{expansion}
C_0+\Delta C =\frac{t_1}{t_0}\frac{\sin((N_1+1)({k}_1(E_0)+{k}'_1(E_0)\Delta E+...))
}{\sin(N_2({k}_2(E_0)+{k}'_2(E_0)\Delta E+...))}\cr\cr
=\frac{t_0}{t_2}
\frac{\sin(N_1({k}_1(E_0)+{k}'_1(E_0)\Delta E+...))}{\sin((N_2+1)({k}_2(E_0)+{k}'_2(E_0)\Delta E+...))}.
\end{eqnarray}
Now, writing $N_i=\nu_iN$ for some integer $N$ we impose the
{\it resonant condition}
$$
\frac{\nu_1{k}'_1(E_0)}{\nu_2{k}'_2(E_0)}=\frac{m_1}{m_2},\quad m_1,m_2\in\NZ,
$$
or in other words
$$m_1=r\,\nu_1{k}'_1(E_0),
\qquad m_2=r\,\nu_2{k}'_2(E_0),\qquad\mbox{for some } r\in\NR,$$
where we have assumed that  $m_1$ and $m_2$ are relative primes.
Then, for large $N$, it is clear that if we take
$$\Delta E=\frac{\pi rn}{N},\quad n\in\NZ,$$
the first subleading terms in the expansion above (for $n=O(1)$)
are
$$N_i{k}_i'(E_0)\Delta E= \pi m_i n$$
and one easily checks that  $E_0+\Delta E$ is another solution of
\pref{main_eq} up to corrections of order $O(1/N)$.
The relevant fact is that these solutions give the same value
for $C_0^2$, and therefore for the leaning, up to $O(1/N)$ terms. This explains
the smooth curves of the Lissajous type that we observe for certain
values of the energy.

To determine the width of the window we must consider the subleading
corrections. They pose a limit to the validity of our approximation.
To be specific let us focus in the argument of the first numerator in
(\ref{expansion}) and write
\begin{eqnarray}
  (N_1+1){k}_1(E_0+\Delta E)&=&
  (N_1+1){k}_1(E_0)+N_1{k}'_1(E_0)\Delta E\cr\cr&+&
{k}'_1(E_0)\Delta E+ \frac12 N_1{k}''_1(E_0)(\Delta E)^2
+\dots
\end{eqnarray}
As we discussed before, given our choice of $\Delta E=\pi r n/N$
the second term of the expansion gives a contribution $\pi m_1n$
and hence, inside the sinus function reduces to a global $\pm1$,
which is the same at both sides of \pref{main_eq} and can be removed.
The next two terms are
$$
\frac{\pi rn}{N}
{k}'_1(E_0)+ 
\frac{\nu_1(\pi r n)^2}{2N} 
{k}''_1(E_0).
$$
Note that for $n=O(\sqrt N)$ the second term above
is of order $O(1)$ and our approximation ceases to be valid.
Hence we conclude that the width of the resonant windows
scales like $1/\sqrt N$.

This is true provided ${k}_1''(E_0)\not=0$ which does not hold
at $E_0=0$. In this case we must go one step further in the
expansion, so that the first corrections are
$$
\frac{\pi r n}{N}
{k}'_1(0)+ \frac{\nu_1(\pi rn)^3}{6N^2}
{k}'''_1(0),$$
which are of order $O(1)$ for $n=O(N^{2/3})$
and the validity of our approach extends as far as
$\Delta E=O(N^{-1/3})$.
This explains why the resonances at the center of the plot, when they
occur, are much wider.

A different question is: how many curves in the $E$-${L}$ plane are there around
a resonant value $E_0$?, or in other words, how many well separated values
for the leaning do we get for values of the energy near $E_0$?
To answer this question we may use the results for the spectral density that
we derived in the previous section.

First, consider that the separation between two consecutive
values for the energy with the same value of the leaning
(up to $O(1/N)$) is $$\delta E=\pi r/N.$$ 
Now, combining this with the spectral density (\ref{frequency}) we
can obtain the number of states between two consecutive repetitions
of the leaning, i. e. the number of curves at the outset of the resonance.
Namely
$$N \lambda_{t_0}(E_0)\delta E=N\frac{\nu_1{k}'_1(E_0)+\nu_2{k}'_2(E_0)}\pi
\frac{\pi r}N =m_1+m_2,$$
where for the last equality the conditions for resonance,
$m_i=r \nu_i{k}'_i(E_0)$, have been used.

\begin{figure}
\hskip -.7cm\includegraphics[width=8.5cm]{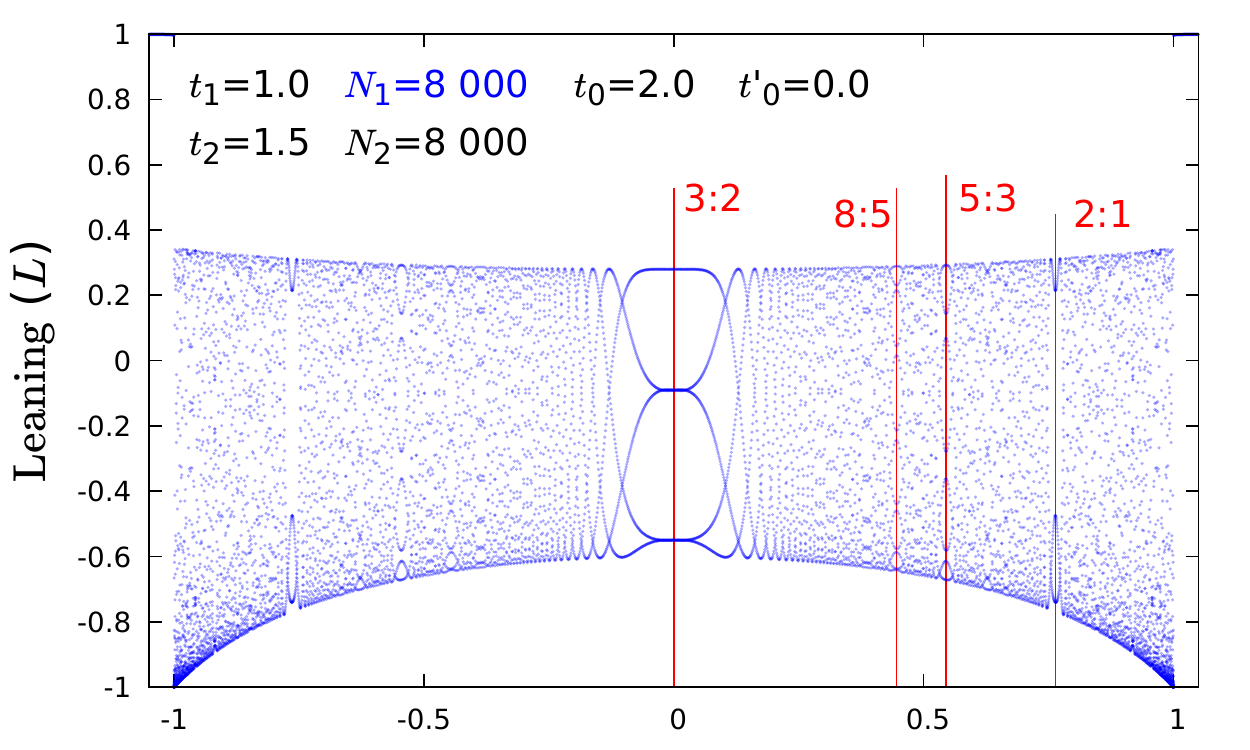}
\hskip -.7cm\includegraphics[width=8.5cm]{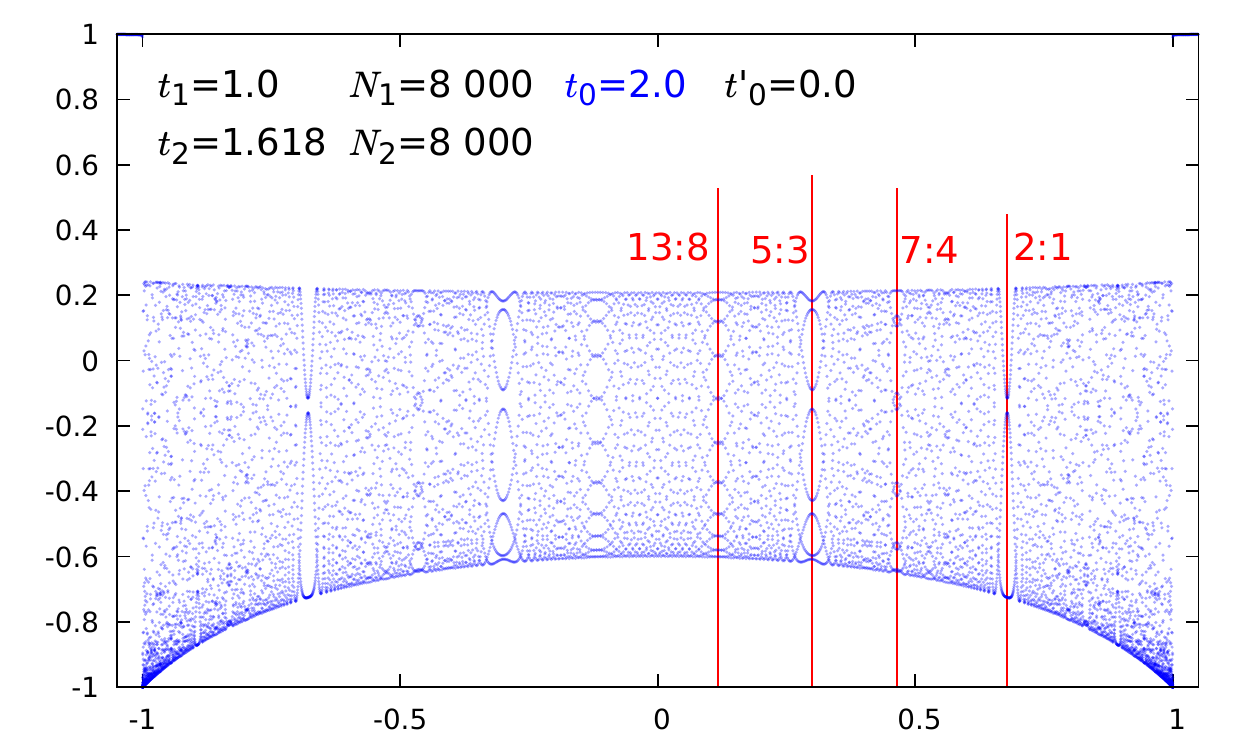}

\hskip -.7cm\includegraphics[width=8.5cm]{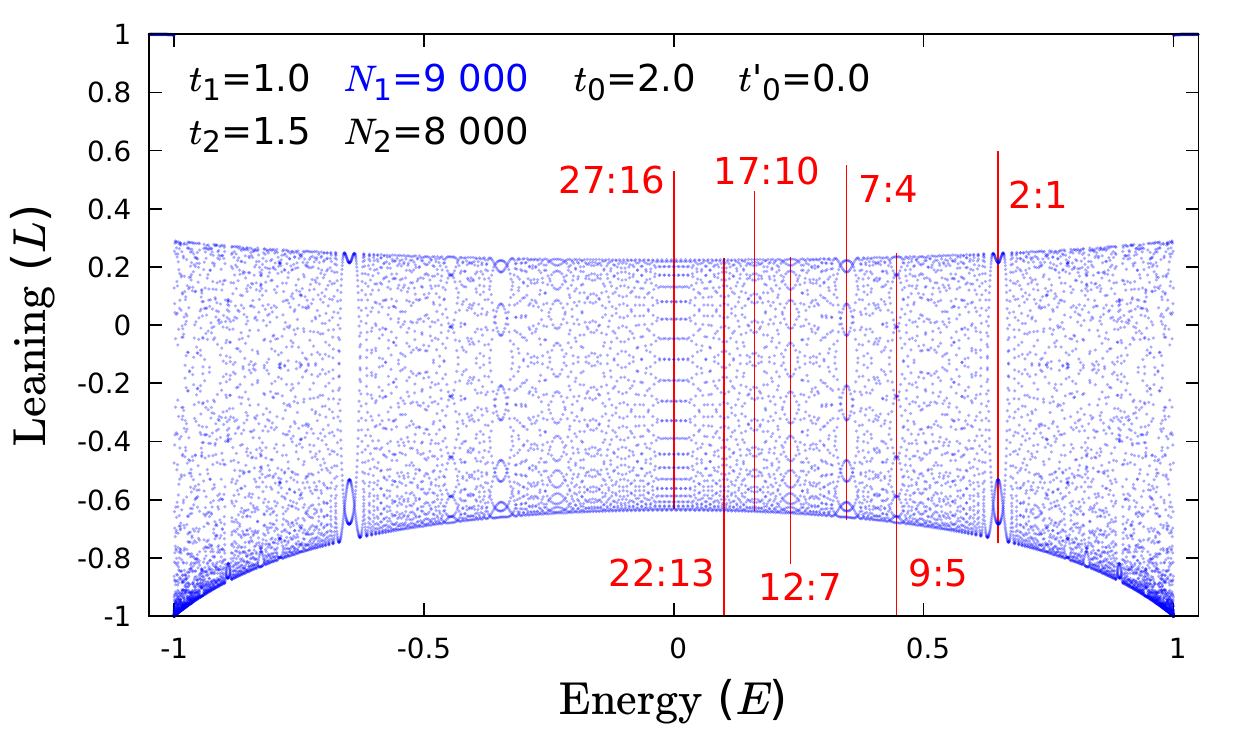}
\hskip -.7cm\includegraphics[width=8.5cm]{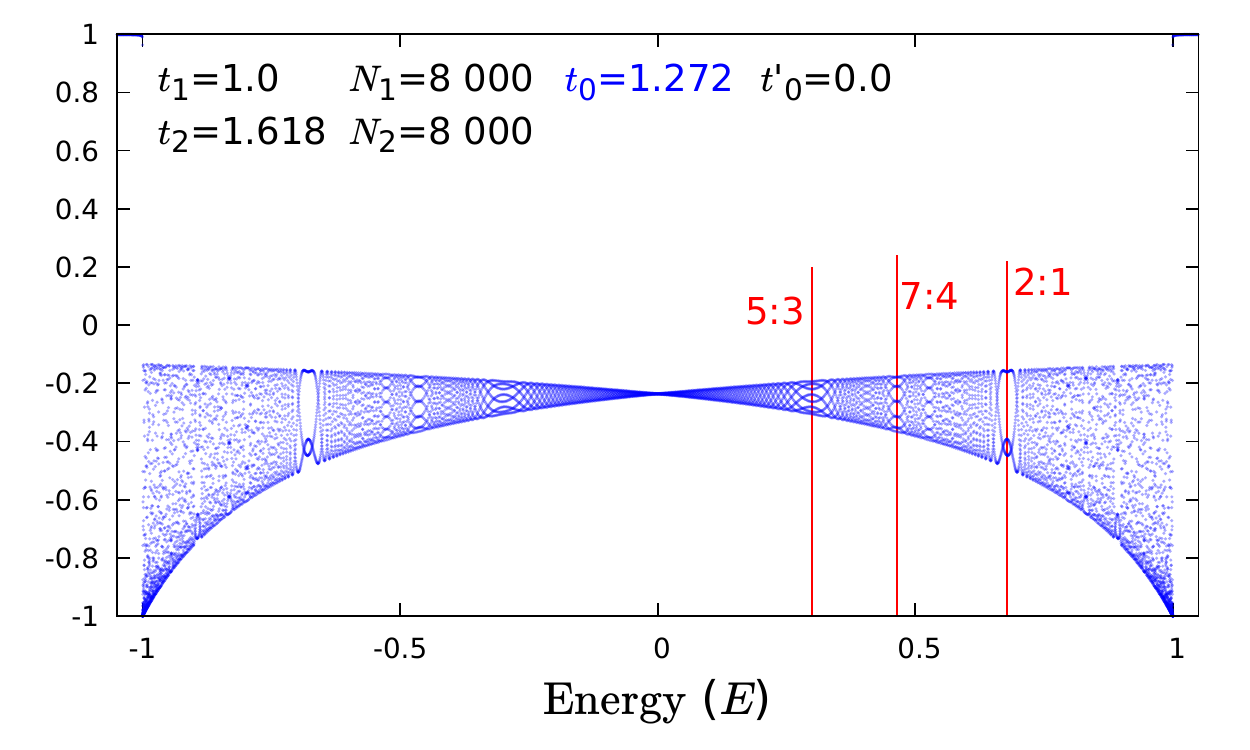}
\caption{In these figures we show the plots of the leaning versus
  energy for different composed systems with the hopping, sizes
  and contacts that appear in every panel. The vertical lines mark the
  value of the energy for which we expect a {\it resonance} according
  to the discussion in section \ref{sec_resonance}.
  The pair of numbers at every vertical line is the theoretical relation
  between the two resonant frequencies.
  Notice that the number of Lissajous type curves at every resonance
  coincides with the sum of those two numbers, as it is explained in the text.}
\label{resonance}
\end{figure}

Then, we conclude that the number of curves that we obtain at the
resonant value is precisely $m_1+m_2$. 
These results are illustrated in fig. \ref{resonance} where we show 
the plot for the leaning and we superimpose some resonant values
obtained according to our derivation. Note that the number of Lissajous
type curves is also correctly predicted.

We must add that at some special points, like for instance at $E_0=0$,
there may appear some degeneracy for the leaning which results in a
smaller number of different values for it. This is clearly observed
in the upper-left panel of fig. (\ref{resonance}), where the crossing of the curves
at $E_0=0$ reduces from five to three the number
of allowed values for the leaning. Of course, this degeneracy occurs only
at  $E_0=0$ and is broken in its vicinity, recovering there the right
number of curves.

As it is clear from the plots and from the discussion in this section,
the resonance windows do not depend on the contact $t_0$ of the two
subsystems but they are sensitive to its size. In the next section
we will discuss a property of the plots that behaves exactly in the opposite
way, i. e. it is independent of the size and varies with the contact
coupling $t_0$.

\section{The boundary.}\label{sec_boundary}

In this section, we will find an analytical
expression for the curves that limit the distribution of
points in the $E$-${L}$ plane.

We are interested in the boundaries of the cloud in the $E$-${L}$ plane
that are valid in the thermodynamic limit, when $N_1,N_2\to\infty$.
To achieve this goal we first look for an upper and lower bound of $C^2$
at a given value of the energy. It is clear that, given the monotonic
decreasing character of the leaning in \pref{leaning} with $C^2$, the 
latter leads respectively to lower and upper bounds for ${L}$.

It also happens that, as we show below, it is possible to obtain
bounds for $C^2$ which are valid for any $N_1,N_2$ and are
optimal, in the sense that they can be approached as much as we want
by varying the size of the two parts of the system.

As we look for bounds for $C^2$ independent of
$N_1, N_2$, it makes sense to replace
$N_1{k}_1(E)$ and $N_2{k}_2(E)$ inside the trigonometric
functions of \pref{main_eq}
by two continuous variables $\xi_1, \xi_2\in[0,2\pi)$ independent,
in principle, of $E$.

To justify this replacement consider, on the one hand side,
that we are looking for bounds for $C^2$,
then if we relax the conditions for the equation
\pref{main_eq} we are sure that the bounds for the
modified equation are still valid for the original one.
On the other hand,
we may argue that by considering
$N_i$ large enough we may approach any value $\xi_i$
as much as we want which implies that our bounds,
valid for any $N_i$, are optimal.

To proceed, we replace the equation \pref{main_eq} by
\begin{equation}\label{bound_eq}
\frac{t_1}{t_0}\frac{\sin(\xi_1+{k}_1(E))}{\sin\xi_2}=\frac{t_0}{t_2}\frac{\sin\xi_1}{\sin(\xi_2+{k}_2(E))},
\end{equation}
and, consequently, 
\begin{equation}\label{cociente}
C^2=\frac{t^2_1}{t^2_0}\frac{\sin^2(\xi_1+{k}_1(E))}{\sin^2\xi_2}.
\end{equation}
If we replace $\xi_i$ by the new variables
$$z_i=\cos{k}_i(E)+\sin{k}_i(E) \cot\xi_i,\quad i=1,2,$$
then equation \pref{bound_eq} is easily solved
$$ z_2 = \frac {t_0}{\bar t_0} z_1^{-1}, $$ 
where for later convenience we have introduced
the {\it dual} contact $\bar t_0= t_1 t_2/t_0$.
Now we use the previous relation
to express $C^2$ in terms of
the single variable $z_1$ to obtain 
\begin{equation}\label{coc_solved}
  C^2=\frac{\sin^2{k}_1}{t_2^2\sin^2{k}_2}
  \frac{\bar t_0^2 z_1 -2 t_0\bar t_0\cos{k}_i+t_0 z_1^{-1}}
  {\bar z_1 -2\cos{k}_i+ z_1^{-1}}.
\end{equation}

Then, we simply have to determine
the maximum and minimum of 
\pref{coc_solved} as a function of $z_1$,
for every value of the energy.
The task is, of course, straightforward but somehow
painful. The final expressions are rather cumbersome
and of little interest to us.
Instead of writing down the analytic expression for
the upper and lower bound of $C^2$ and the leaning, we prefer
to plot it for some special cases.

Notice that, as the bounds for $C^2$ are independent of $N_1,N_2$,
those of ${L}$ only depend on $\nu_1$ and $\nu_2$, namely
$${L}=\frac{\nu_1C^2-\nu_2}{\nu_1C^2+\nu_2}.$$

\begin{figure}
\hskip -.7cm\includegraphics[width=8.5cm]{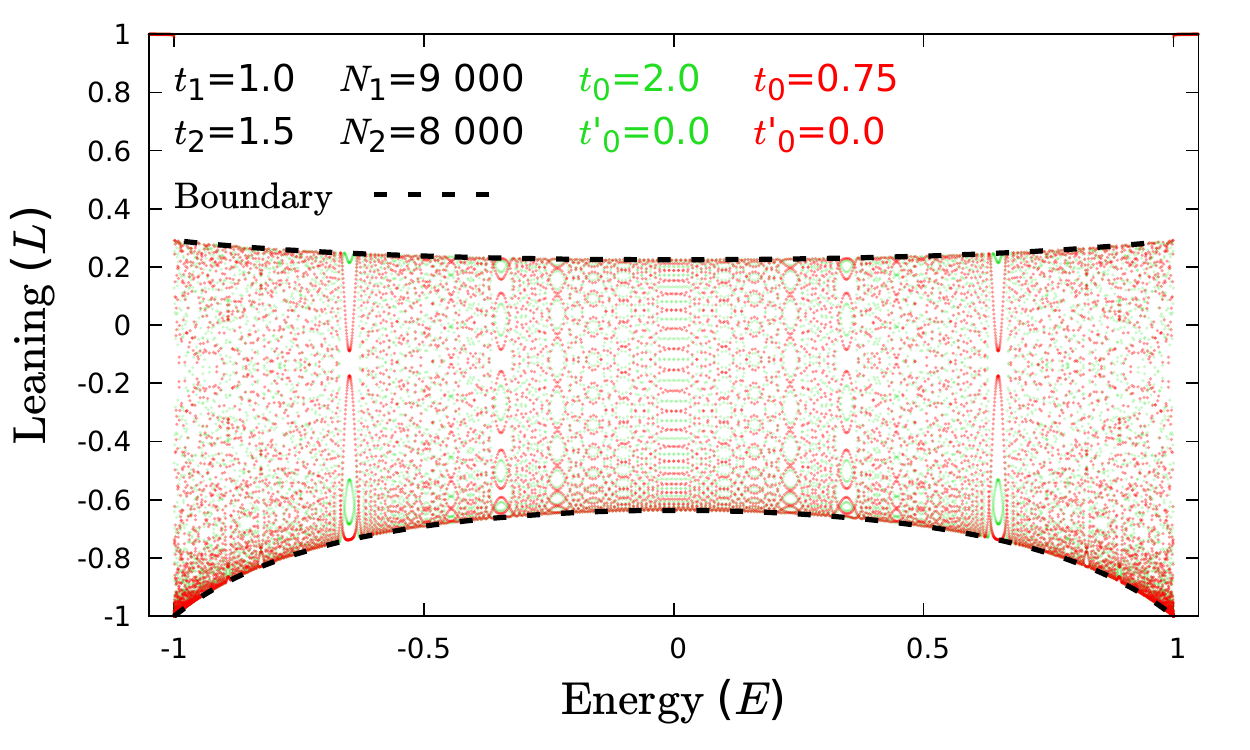}
\hskip -.7cm\includegraphics[width=8.5cm]{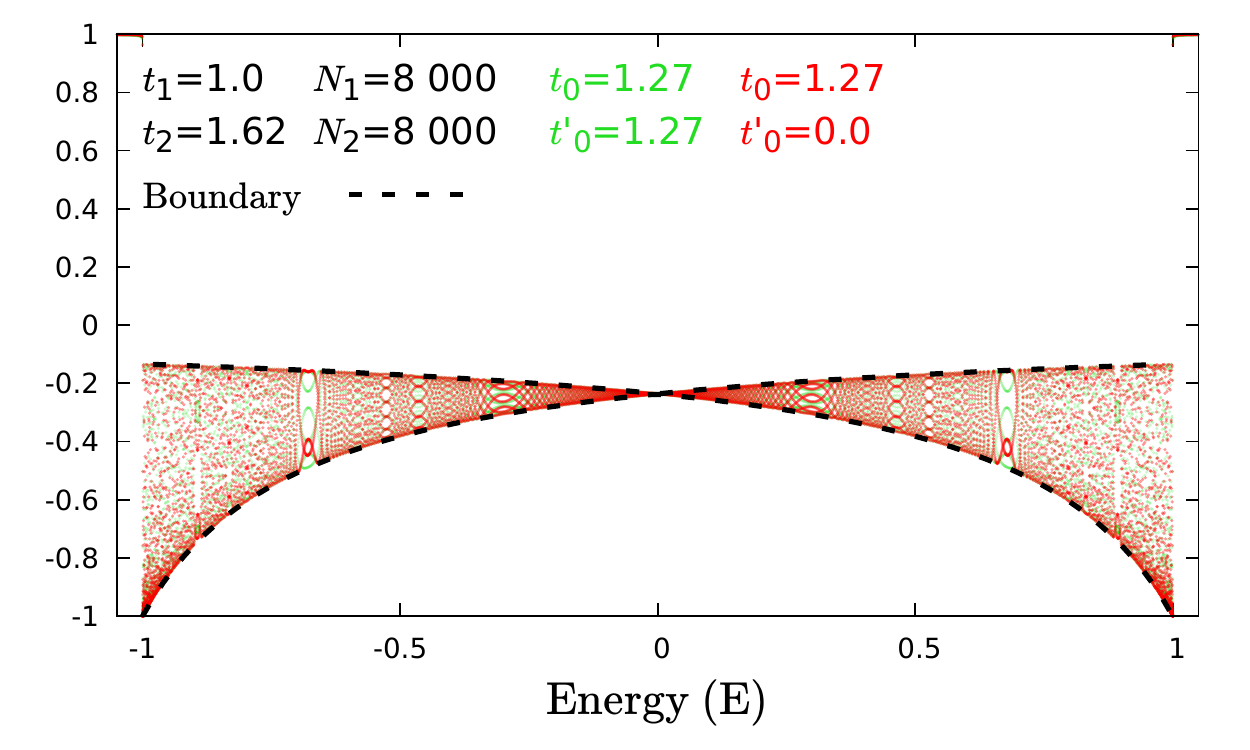}
\caption{In these plots we show the distribution for the leaning
  versus the energy of different chains. In the left panel the two systems
  are related by duality $t_0\leftrightarrow t_1t_2/t_0$.
  It is clear that although the two distributions of points
  (represented in different colors) are not identical
  they fill the same region. In the panel on the right we represent the selfdual case $t_0=\sqrt{t_1t_2}$ and the two distributions differ by the {\it largest}
  contact $t'_0$. The situation is similar to the previous one, different distribution but the same boundaries. The discontinuous lines represent the theoretical boundaries obtained as explained in the text.}
\label{boundary}
\end{figure}

The comparison between the analytic and numerical results
for different relative sizes and values of the contact
are collected in the plots of fig. \ref{boundary}.

An interesting fact that we would like to emphasize
is the duality between higher and lower
contact. In fact, we can show that the bounds are
unchanged if we replace $t_0$ by
$$\bar t_0=t_1t_2/t_0.$$
A duality that maps zero
to infinity or, as we mentioned above, higher to lower
coupling constant. 

The duality is easily proven starting from \pref{coc_solved}.
There we see that the value of $C^2$ is unchanged if we replace
$t_0$ by $\bar t_0$ and simultaneously $z_1$ by $z_1^{-1}$. Therefore,
the maximum and minimum for $C^2$ are unchanged under the duality.

In the left panel of fig. \ref{boundary}, we plot the leaning for
dual values of the contact
and we see that the respective allowed regions
perfectly match. On the right side of 
this figure, the leaning for a self dual value of the 
contact $t_0=\sqrt{t_1t_2}$ is plotted.

Let us discuss now which is the effect of taking $t'_0\not=0$.
First notice that if we replace in \pref{gluing} $\delta_i$ by 
$\delta_i-N_i{k}_i(E)/2,\ i=1,2$, which is a simple redefinition 
of the unknowns, we obtain a more symmetric equation. Actually, 
replacing now $\delta_i+N_i{k}_i/2$ by $\xi_i$ and $N_i{k}_i/2-\delta_i-{k}_i$ 
by $\eta_i$ we obtain the equivalent equations:
\begin{eqnarray}
C=\frac{t_1}{t_0}\frac{\sin(\xi_1+{k}_1(E))}{\sin\xi_2}=\frac{t_0}{t_2}\frac{\sin\xi_1}{\sin(\xi_2+{k}_2(E))},\cr
C=\frac{t_1}{t'_0}\frac{\sin(\eta_1+{k}_1(E))}{\sin\eta_2}=\frac{t'_0}{t_2}\frac{\sin\eta_1}{\sin(\eta_2+{k}_2(E))}.
\end{eqnarray}
Notice that the second line is like the first one by simply replacing
$\xi$ by $\eta$ and $t_0$ by $t_0'$. Then we have two equations for
obtaining bounds on $C^2$, one with $t_0$ and another with $t_0'$.
It happens that the smaller $t_0+\overline t_0$ is
(its minimum value is attained for the selfdual case $t_0=\sqrt{t_1t_2}$)
the more restrictive the bounds are,
and this applies both for the upper and the lower bound.

Consequently, only one of the contacts matters for determining
the boundaries of the allowed region in the $E$-${L}$ plane, namely
the one closer to the selfdual value or equivalently
the one with the least value for $t_0+\overline t_0$.

More formally, if we introduce an order relation defined by:
$t_0\prec t_0'$ if and only if
$t_0+\overline {t_0} < t'_0+\overline {t'_0}$, the {\it smallest} of the two
contacts determines the shape of the cloud.

This can be checked in the numerical experiments 
where it is apparent that a modification of the
{\it larger} contact does not alter the shape of the cloud,
as can actually be seen in the right plot of fig. \ref{boundary}.  

\section{The probability measure}\label{sec_measure}

If we examine the different plots of previous sections, one observes
that for most of the allowed region the points that represent
$(E,{L})$ pairs form a cloud more dense near the boundaries and more sparse
at the middle. Of course, the previous is not true at the resonance
windows, where the points form definite curves. But, as we discussed
in section \ref{sec_resonance}, one can prove that the resonant regions shrink
with the size of the system and eventually disappear in the
thermodynamic limit.

Then the question that might have sense and we will study is
whether there is a measure in the $E$-${L}$ plane 
that represents the density of points in the
thermodynamic limit and how it depends on the parameters of the system.
We believe that such a measure exists and for $t_0,t'_0\not=0$
is absolutely continuous with respect to the Lebesgue measure.

To be more precise, take $N_1=\nu_1 N$ and $N_2=\nu_2 N$,
let the hopping parameters be as usual
$t_2>t_1>0$ and contacts $t_0,t_0'$. Now
we define the following probability measure on the Borelians
$S\subset X=[-t_1,t_1]\times[-1,1]$
$$
\mu_N(S) = K_N \, \sharp\{\Psi_{E}\,|\,(E,L_{E})\in S\},
$$
where by $\sharp$ we denote the cardinality of the set,
$\Psi_{E}$ is an eigenfunction of the Hamiltonian of the composite 
system with chains of length  $\nu_1N$ and $\nu_2N$,
and ${L}_{E}$ is the leaning of $\Psi_E$.
$K_N$ is the appropriate normalization constant to obtain a
probability measure, i. e. $\mu_N(X)=1$.

We assert that these measures converge, when $N\to\infty$,
to a probability measure $\mu$ on the Borelians of $X$.

\begin{figure}
\center
  \includegraphics[width=14cm]{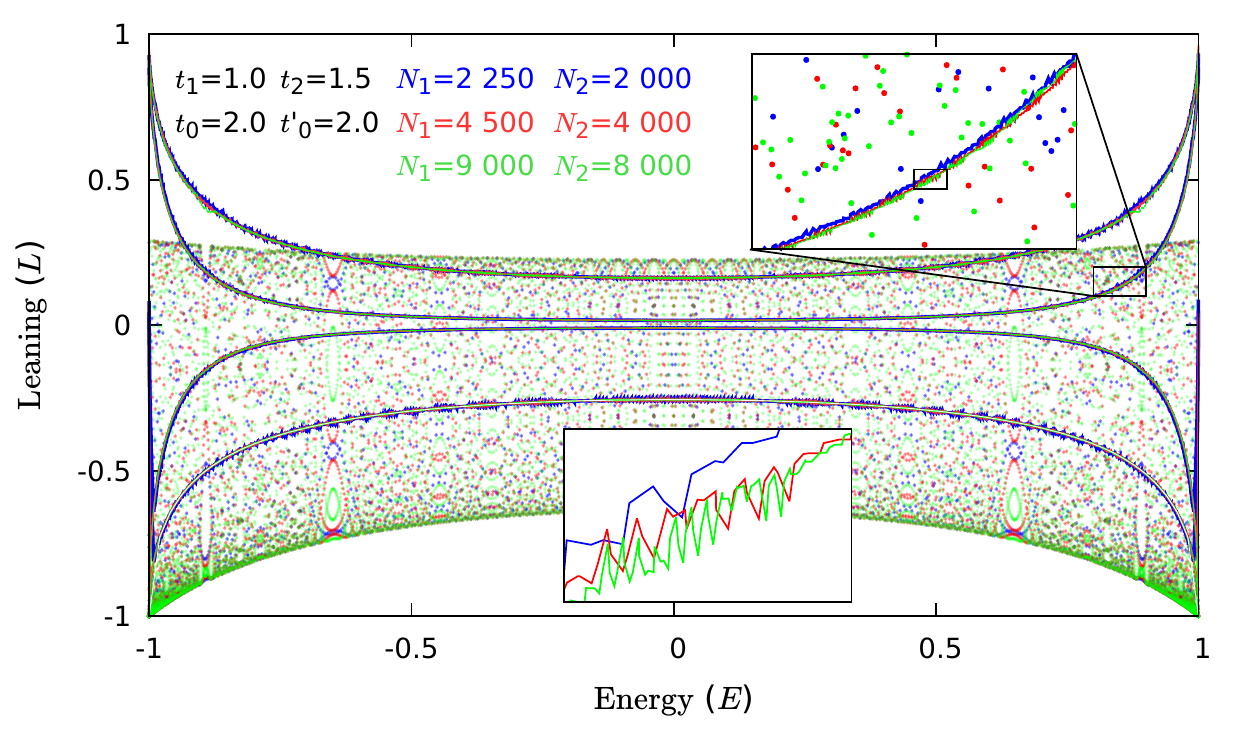}
  \caption{In this plot we represent the distribution of points in the
    $L-E$ plane for chains that differ in size, while sharing the proportion
    between the two subsystems. Every size is represented in a different color.
    The curves are the running average (over 200 points) of $L^n$ for
    $n=1, 7, 6, 2$ respectively from bottom to top. It is clear that
    the curves for different size almost coincide
    and the two consecutive magnification insets indicate that a well defined
    limit for the expectation value of all momenta, when $N\to\infty$, seems
    to exist.}
\label{moments_diff_N}
\end{figure}

We do not have a proof for the existence of
$\mu$, only numerical evidences based on the good behaviour of
the expected value for different random variables and its apparent convergence
with $N$ as illustrated in fig. \ref{moments_diff_N}.
There we show the plot of the leaning for different values of $N_1$ and $N_2$
with the same relative sizes $N_1/N_2$ and the running average $\langle L^n\rangle$
for $n=1$ that corresponds to the lowest (blurry) curve
and going upwards $n=7, 6, 2$. We see that the curves for
different sizes of the system agree to a large extent and they seem to
have a smooth large $N$ limit.
The well defined limit for the different moments
is a strong numerical indication of the 
existence of a Borelian measure when $N\to\infty$.

Sometimes it will be important to emphasize the dependence of the limiting
measure on some of the parameters of the theory, in that case we will write
those parameters as subindices. In the following, we will focus mainly on
the dependence of $\mu_N$ or $\mu$ on the contacts, so we will write it
$\mu_{N,t_0,t'_0}$ or $\mu_{t_0,t'_0}$.
The first observation is that due to the parity invariance of the Hamiltonian
$\mu_{N,t_0,t'_0}=\mu_{N,t'_0,t_0}$; therefore, if the limit exists
we must have $\mu_{t_0,t'_0}=\mu_{t'_0,t_0}$.

In general, we do not know how to determine $\mu_{t_0,t_0'}$, only
when $t_0=t'_0=0$ (or $\infty$) because, in this case,
the chain splits into two independent
homogeneous systems with well defined leaning.
Therefore, $\mu_{0,0}$ can be obtained as the normalized
sum of the corresponding spectral measures. Hence, we have
$${\rm d}\mu_{0,0}=\big(\rho_1(E)\delta({L+1})+\rho_2(E)\delta(L-1)\big) {\rm d}E\, {\rm d}L,$$
where
$$\rho_i(E)=\frac{\nu_i/\sqrt{t_i^2-E^2}}{\pi\nu_1+2\nu_2\arcsin(t_1/t_2)},$$
and $\delta$ represents the Dirac delta function. That is, in this case
the measure is supported in the upper and lower boundary of $X$,
with $L=\pm1$.

As we said, except for this trivial case, we are not able to determine $\mu$.
However, based on numerical experiments and some
analytical hints we can establish some conjectures that
we present now.

\begin{figure}
\center
  \includegraphics[width=14cm]{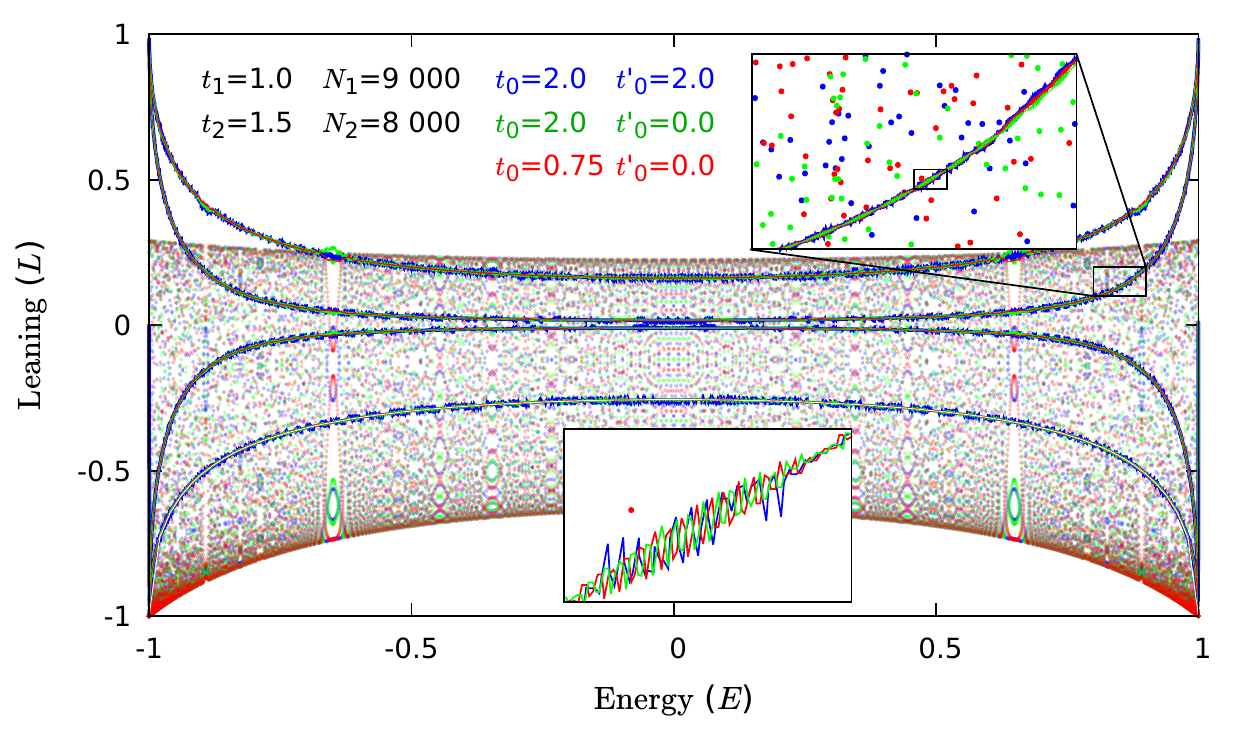}
  \caption{In this plot we show the distribution of points and
    running average of the momenta $L^n$ ($n=1,7,6,2$ respectively for
    the curves from bottom to top) for three
    different chains. They differ by the value of $t'_0=2.0$ or $0.0$
    and/or by the replacement of $t_0=2 $ by its dual value
    $t_1t_2/t_0=0.75$. We see that in this case the running average
    of the three different chains coincide. This coincidence is more
    striking if we look at the two consecutive magnifying insets.} 
\label{moments_diff_t0p}
\end{figure}

\begin{itemize}
  
\item[{\bf 1.}] According to the discussion of section \ref{sec_boundary},
the support of the measure is in the region between the curves
$L_{\rm max}(E), L_{\rm min}(E)$. Moreover, we proved in that section that the
limiting curves depend only on the {\it smallest} contact
i. e. they depend only on $t_0$ provided $t_0\prec t'_0$. 
\hfill\break
The stronger conjecture, suggested by different numerical experiments
shown in fig. \ref{moments_diff_t0p}, is that not only the
support of the
measure  depends only on the {\it lowest} contact, but the measure itself
has the same property i. e. we conjecture
$$\mu_{t_0,t_0'}= \mu_{t_0,t_0''},\quad {\rm if}\ t_0\prec t'_0, t''_0.$$
For this reason, and in order to simplify the notation, from now on we will
refer to the measure writing only the {\it smallest} contact $\mu_{t_0}$.

\item[{\bf 2.}] Another property of the limiting curves that we showed in
  section \ref{sec_boundary} is its invariance under duality
  $t_0\mapsto \bar{t}_0=t_1t_2/t_0$. Based again in numerical evidences,
  see fig. \ref{moments_diff_t0p} for an example, we conjecture
  $$\mu_{t_0}=\mu_{{\bar{t}_0}}.$$
  That is, we assert that not only the support of the measure is left
  invariant under the duality transformation, but also the measure itself.
\end{itemize}

\begin{figure}
\center
  \includegraphics[width=14cm]{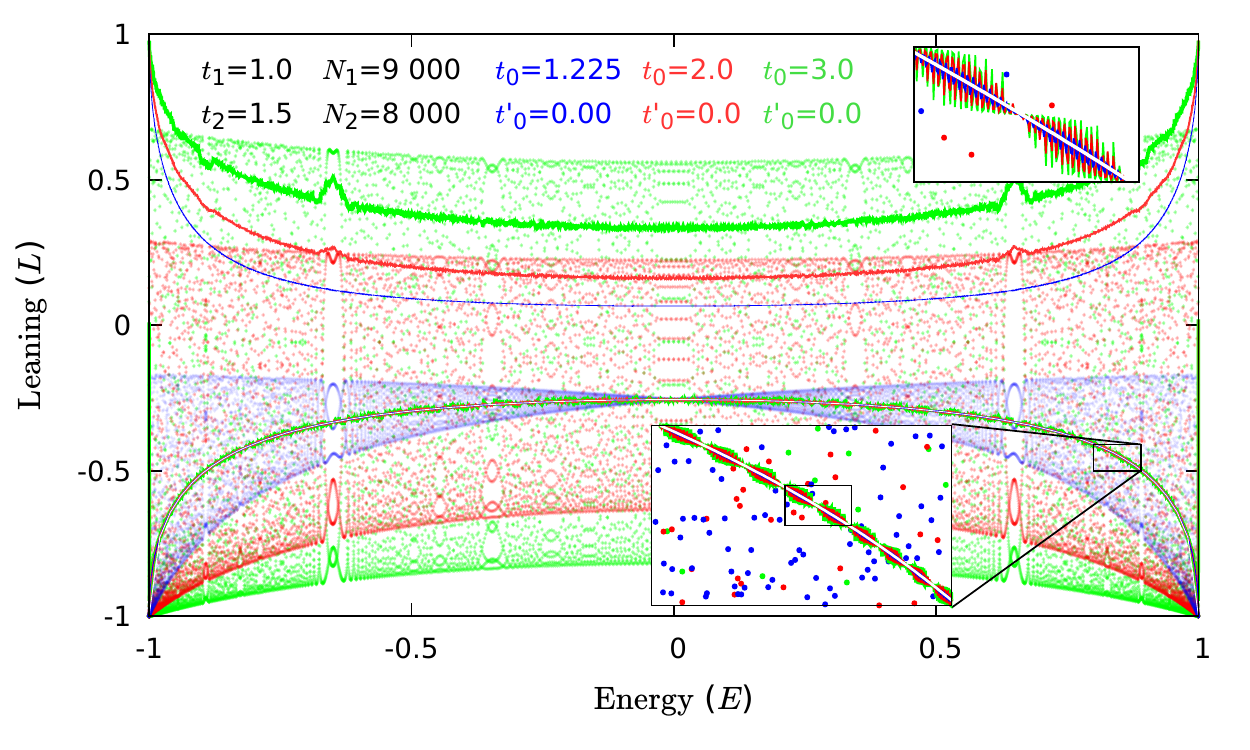}
  \caption{In this plot we show the distribution of points and
    running average $\langle L\rangle$ and
    $\langle L^2\rangle$,  over 200 points, for
    chains that differ in the contact, $t_0$, and are represented in
    different colors.
    While the three curves corresponding to $\langle L^2\rangle$ are
    clearly different, those for $\langle L\rangle$ coincide as it is made
    manifest in the insets. The curve in white (visible in the insets)
    represents the predicted value. The explanation for this fact is presented
    in the appendix.}
\label{moments_diff_t0}
\end{figure}

  We do not have any analytic argument to substantiate these two last
properties. But if we compute the running moments in ${L}$ of the
distributions, when varying $t'_0$ or when replacing $t_0$ by
$\overline t_0$, we find that they are as close as
they possibly could be. This is shown in fig. \ref{moments_diff_t0p}.
\begin{itemize}

\item[{\bf 3.}] From the dependence of the clouds with $t_0$, illustrated
  in fig. \ref{moments_diff_t0}, it seems reasonable to conjecture that, 
  except for $t_0=t'_0=0$ (or $\infty$), the measure is absolutely
  continuous with respect to the Lebesgue measure. That is, there is a
  function $\sigma_{t_0}\in L^1(X)$ such that
  $${\rm d}\mu_{t_0}=\sigma_{t_0}(E,L)\,{\rm d}E\, {\rm d}L.$$

\item[{\bf 4.}] Although we can not determine $\mu_{t_0}$,
  based on the arguments of section \ref{sec_spectrum}
  we believe that the marginal distribution
  for $E$ does not depend on $t_0$.
  Therefore, we can write
   $$\displaystyle \hat{\lambda}_{t_0}(E)\equiv\int_{-1}^1\sigma_{t_0}(E,{L}) {\rm d}{L}=
  \frac{\nu_2/{\sqrt{t_2^2-E^2}}+\nu_1/{\sqrt{t_1^2-E^2}}}
       {2\nu_2\arcsin(t_1/t_2)+\nu_1\pi},$$
       where the right hand side has been computed using the spectral density
       obtained in section \ref{sec_spectrum} or, alternatively, the measure
 that we determined before for $t_0=0$. 
 
\item[{\bf 5.}] Finally, we can prove that the expected value of $L$
  at fixed value of the energy is also
  independent of $t_0$. Indeed, using again the probability measure at
  $t_0=0$, we have
   \begin{equation}\label{running_avg}
\displaystyle \langle
{L}\rangle_E\equiv\hat\lambda_{t_0}(E)^{-1}\int_{-1}^1\sigma_{t_0}(E,{L})
{L} {\rm d}{L} = \frac{\nu_2/{\sqrt{t_2^2-E^2}}-{\nu_1}/{\sqrt{t_1^2-E^2}}}
        {\nu_2/{\sqrt{t_2^2-E^2}}+{\nu_1}/{\sqrt{t_1^2-E^2}}}.
\end{equation}
       As it is explained in the appendix, this result can be proven
       by estimating the running average of $L$ in the large $N$ limit.
       Numerical experiments also support our result.
       These are shown in fig. \ref{moments_diff_t0}
       where we present the running average of the leaning
       that we obtained numerically for different values of $t_0$
       (the {\it thick} curve
       transversing the cloud in its lower part), and we check that it
       is independent of $t_0$
       and agrees extremely well with the conjectured predictions.
       If we look at the two magnifying insets it is clear that
       curves for different $t_0$, represented in different colors,
       agree perfectly. They also coincide with the theoretical value
       of (\ref{running_avg}) that we plot in white and is the line
       that cuts right in the middle the numerical curves.
       In contrast, the numerical curves for $\langle L^2 \rangle$ for
       different values of $t_0$ (in different colors, in the upper part
       of the plot) are well separated.
\end{itemize}

\section{Conclusions, generalizations and extensions}\label{sec_conclu}

We have shown that the probability of presence for one particle in a
composite system, characterized here with the leaning, follows a rather
intricate pattern.

We have been able to unravel many of its properties, like the nature of the
resonant regions, the boundaries of the allowed region, its independence of
the {\it largest} contact, the duality between large and small contact coupling
or the universal properties of the average leaning.

It is interesting to remark that the leaning, in the thermodynamic limit, is
definitely a non perturbative property. This can be argued in several ways:
first, the duality discussed in sections \ref{sec_boundary}
and \ref{sec_measure} and mentioned above allows to identify the small and large coupling
constant region $t_0\leftrightarrow t_1t_2/t_0$; second, we observe that for
no matter how little $t_0\not=0$ is we may find (for $N$ large enough)
states with a leaning arbitrary close to $0$, very far from the
unperturbed system where $|L|=1$ for any state.

Suggestively enough, the non perturbative character of the
leaning can be traced back to the existence (for large $N$) of {\it small
denominators} in the quantum perturbative expansion.
This is reminiscent of the same phenomena in the canonical perturbation
expansion in classical mechanics, which is one of the essential ingredients 
for the existence of non integrable systems and chaotic dynamics.

Here, of course,  we may not have
{\it sensitive dependence on initial conditions} for the evolution, as the
dynamics is linear;
but, instead, the expected position of one particle stationary states depends
sensitively on its energy.

While, as we just stated, the leaning depends critically on the energy
of the stationary state,
we may obtain a predictable result if we consider the average over a
range of energy.
This is observed numerically and can be rigorously proven. 
For the latter proof we have to get
rid of the small denominators problem and it is
interesting to remark that the way we proceed is very much reminiscent
of the analogous strategy for the proof of KAM theorem in classical perturbation
theory: we fix initially a cut-off that suppresses the small denominators,
then we can proceed with the different estimates before removing
the cut-off.

An open problem is to compute the density in the $E$-$L$ plane for
the stationary states in the thermodynamic limit.
There are strong numerical indications that such a Borelian measure exists,
it is absolutely continuous with respect to the Lebesgue measure
(except for $t_0$ and $t_0'$ equal to $0$ or $\infty$)
and varies continuously with $t_0$ in the total variation topology
(the convergence for  $t_0, t_0'\to 0,\infty$ should occur
only in the weak topology).

Let us remark that the previous results were obtained for
the simplest kind of systems composed of two homogeneous
tight-binding chains joined at every end by links
with different hopping parameter.
However, the behaviour that we have described in the paper
seems rather universal and it has been
observed for the SSH chain (alternating hopping), for the Ising chain
and, more generally, for the $XY$ spin chain.
One can consider also different types of contacts (of finite range)
without affecting the essential properties of the leaning.
Particular examples beyond tight binding models will be presented elsewhere.

Another interesting issue connected to this paper is the continuum limit
of our system (different from the thermodynamic limit that we work
out here). The natural guess is a free fermion in one dimension with
boundary conditions, impurities and/or localized (delta like) potentials.
At present we do not know if a similar pattern for the leaning in this kind
of systems occurs. This is a very interesting question and it could
help to clarify the weak/strong coupling duality that we uncover in this work.
In fact, for a free fermion, the latter can be explained by
the existence of dualities in the bosonization process. A similar
explanation could also be valid here.

Finally, we would like to comment that all the systems mentioned
so far can be mapped to free fermionic chains
and therefore can be analyzed with relatively ease.
It would be interesting to go beyond that and explore the behaviour of the
leaning for systems composed of interacting chains like the Hubbard model
or others. We plan to approach these problems in our future research.
\newline 

\textbf{Acknowledgments:} Research partially supported by grants E21 17R, DGIID-
DGA, and PGC2018-095328-B-100, MINECO (Spain). FA is supported by Brazilian
Ministries MEC and  MCTIC and acknowledges the warm hospitality and
support of Departamento de F\'{\i}sica Te\'orica, Universidad de Zaragoza,
during several stages of this work.
We would like to acknowledge one of the referees for interesting remarks
concerning the continuum limit.

\appendix
\section{}

In this appendix, we present a proof of the
invariance of the leaning average under
a change of the contact.

More precisely, given a fermionic chain like the one described in section
\ref{sec_basic} ($0<t_1<t_2$ and, for simplicity, $t_0'=0$) we compute the running
average of the leaning in an interval of energy
$[E-\Delta E,E+\Delta E]\subset(-t_1, t_1)$.

For that, let us denote the spectra for contact $t_0$ and $0$ respectively
by $\Sigma_{t_0,N}=\{E_m; m\in M\}$ and
$\Sigma_{0,N}=\{E^0_{\widetilde m}; \widetilde m\in M_0\}$
as before, and the normalized eigenstates
by $\psi_m$ and $\varphi_{\widetilde m}$ respectively.
Introduce $R\subset M$ such that
$$\{E_m; m\in R\}=
  \Sigma_{t_0,N}\cap[E-\Delta E, E+\Delta E],$$
  and similarly $R_0\subset M_0$ for
  $\Sigma_{0,N}\cap[E-\Delta E, E+\Delta E]$. With these data we define the density matrices
$$\rho=\frac1{\sharp\,(R)}\sum_{m\in R}\ket{\psi_m}\bra{\psi_m},$$
and
$$\rho_0=\frac1{\sharp\,(R_0)}\sum_{\widetilde m\in R_0}\ket{\varphi_{\widetilde m}}\bra{\varphi_{\widetilde m}}.$$

We will prove that in the thermodynamic limit
$$\lim_{N\to\infty}{\rm Tr}\left((\rho-\rho_0)(P_2-P_1)\right)=0.$$
Or in other words, the leaning averaged over a range of energy
does not depend on the contact in the thermodynamic limit.

To show it we express $\rho$ in terms of the unperturbed basis,
$$\rho=\frac1{\sharp\,(R)}
\sum_{r\in R} \sum_{\widetilde m,\widetilde m'\in M_0}
U_{\widetilde m,r}\overline U_{\widetilde m',r}
\ket{\varphi_{\widetilde m}}\bra{\varphi_{\widetilde m'}},
$$
where $U$ is the unitary matrix corresponding to the change of basis,
that is $U_{\widetilde m,r}=\prod{\varphi_{\widetilde m}}{\psi_r}$.

Now we decompose the sets of indices into two disjoint sets,
$M=R\cup P$ and $M_0=R_0\cup P_0$, and write
\begin{eqnarray}
  \rho&=&\frac1{\sharp\,(R)}\sum_{r\in R}
  \left(
  \sum_{\widetilde r,\widetilde r'\in R_0}
  U_{\widetilde r,r}\overline U_{\widetilde r',r}
  \ket{\varphi_{\widetilde r}}\bra{\varphi_{\widetilde r'}}+
  \sum_{\widetilde r\in R_0,\widetilde p\in P_0}
  U_{\widetilde r,r}\overline U_{\widetilde p,r}
  \ket{\varphi_{\widetilde r}}\bra{\varphi_{\widetilde p}}+
  \right.
  \cr
&&\hskip 1.8cm\left.\sum_{\widetilde r\in R_0,\widetilde p\in P_0}
  U_{\widetilde p,r}\overline U_{\widetilde r,r}
  \ket{\varphi_{\widetilde p}}\bra{\varphi_{\widetilde r}}+
  \sum_{\widetilde p,\widetilde p'\in P_0}
  U_{\widetilde p,r}\overline U_{\widetilde p',r}
  \ket{\varphi_{\widetilde p}}\bra{\varphi_{\widetilde p'}}
  \right).
\end{eqnarray}

Due to the orthonormality properties of $U$ we can rewrite the first
term as
$$\sum_{r\in R}
  \sum_{\widetilde r,\widetilde r'\in R_0}
  U_{\widetilde r,r}\overline U_{\widetilde r',r}
  \ket{\varphi_{\widetilde r}}\bra{\varphi_{\widetilde r'}}=
  \sum_{\widetilde r\in R_0}
  \ket{\varphi_{\widetilde r}}\bra{\varphi_{\widetilde r}}-
\sum_{p\in P}
  \sum_{\widetilde r,\widetilde r'\in R_0}
  U_{\widetilde r,p}\overline U_{\widetilde r',p}
  \ket{\varphi_{\widetilde r}}\bra{\varphi_{\widetilde r'}}.
  $$

  With the above replacement and taking into account that
  the stationary states of the unperturbed Hamiltonian are also eigenstates
  of the leaning operator:
  $$(P_2-P_1)\varphi_{\widetilde m}
  =\epsilon_{\widetilde m}\,\varphi_{\widetilde m}
  \ \mbox{ with }\ \epsilon_{\widetilde m}=\pm1,$$
  we can write 
  \begin{eqnarray}\label{rhorho0}
    \Tr(\rho(P_2-P_1))&=&\frac{\sharp(R_0)}{\sharp(R)}
\Tr(\rho_0(P_2-P_1))\cr
&&
+\frac{1}{\sharp(R)}\left(
  \sum_{\widetilde p\in P_0,r\in R} \epsilon_{\widetilde p} |U_{\widetilde p r}|^2
    -\sum_{\widetilde r\in R_0,p\in P} \epsilon_{\widetilde r} | U_{\widetilde r p}|^2
    \right).
  \end{eqnarray}

  Now we must estimate the matrix elements of $U$.
  This can be done by means of the identity
  \begin{equation}\label{matrixelement}
    |U_{\widetilde m m}|^2=\frac{|\bra{\varphi_{\widetilde m}}H_I\ket{\psi_{m}}|^2}
    {(E^0_{\widetilde m}-E_m)^2},
    \end{equation}
    where we can take advantage of the fact that $\varphi_{\widetilde m}$ and
    $\psi_{m}$ are extended wave functions and $H_I$ acts only locally
    at the interface of the two components of the chain. Indeed, one has
   $$ |\bra{\varphi_{\widetilde m}}H_I\ket{\psi_{m}}|=O(N^{-1}).$$

    The problem, however, is that the denominator
    $(E^0_{\widetilde m}-E_m)^2$ for large $N$ and particular values of $m$
    and $\widetilde m$ can be very small (even smaller than $1/N^2$).
    This is an instance of the small denominator problem in
    quantum mechanics.

    To avoid this potential divergence  we must introduce a cut-off. A similar
    strategy (although much simpler in this case) to the one followed
    for proving the KAM theorem in classical mechanics. 

    For instance, to estimate the second term in (\ref{rhorho0}), 
    $$|F|\equiv \frac{1}{\sharp(R)}
      \bigg|\sum_{\widetilde p\in P_0,r\in R} \epsilon_{\widetilde p} |U_{\widetilde p r}|^2\bigg|
      \leq
       \frac{1}{\sharp(R)} \sum_{\widetilde p\in P_0,r\in R} |U_{\widetilde p r}|^2,
       $$
       we fix a range of energy $\delta E<\Delta E$
       and decompose the set $R$ into two disjoint subsets
       $R=R_<\cup R_>$ with
       $$R_<=\{r\in R \mbox{ s. t. } |E_r-E|<\Delta E-\delta E\},$$
       which implies that $|E_{\widetilde p}-E_{r_{_<}}|>\delta E$, for any
       $\widetilde p\in P_0, r_{_<}\in R_<$, and the small denominator problem
       is relegated to the set of indices $R_>$.

       Thus we have
       \begin{eqnarray}\label{estimate1}
         |F|&\leq& \frac{1}{\sharp(R)}
         \left(\sum_{\widetilde p\in P_0,r_{_>}\in R_>} |U_{\widetilde p r_{_>}}|^2
         + \sum_{\widetilde p\in P_0,r_{_<}\in R_{_<}} |U_{\widetilde p r_{_<}}|^2\right)
         \cr &\leq&  \frac{\sharp(R_>)}{\sharp(R)}
         +  \frac{1}{\sharp(R)} \sum_{\widetilde p\in P_0,r_{_<}\in R_<}\frac{|\bra{\varphi_{\widetilde p}}H_I\ket{\psi_{r_{_<}}}|^2}
  {(\delta E)^2},
\end{eqnarray}
       where we have used the normalization condition for the rows of $U$
       to estimate the first term and (\ref{matrixelement}) together with
       the bound for the energy difference for the second.

       Now
       $$\sum_{\widetilde p\in P_0}|\bra{\varphi_{\widetilde p}}H_I\ket{\psi_{r_{_<}}}|^2\leq
       \bra{\psi_{r_{_<}}}H_I^2\ket{\psi_{r_{_<}}}\leq \frac{2t_0^2}{M},
       $$
       with
       $$M=\min\left\{\nu_1 N-\frac{t_1^2}{\sqrt{t_1^2-(|E|+\Delta E)^2}},
       \nu_2 N-\frac{t_2^2}{\sqrt{t_2^2-(|E|+\Delta E)^2}}\right\}.$$
       The important fact here is that for fixed
       $|E|+\Delta E< t_1< t_2$ we have $M=O(N)$ for large $N$.
       
       Inserting this into (\ref{estimate1}) and performing the sum,
       we obtain
       $$|F|\leq \frac{\sharp(R_>)}{\sharp(R)} + \frac{2t_0^2}{M(\delta E)^2}.$$

       The same estimate can be used for the third term on the right
       hand side of (\ref{rhorho0}) to get
       \begin{equation}\label{mainuneq}
       |\Tr((\rho-\rho_0)(P_2-P_1))|\leq\frac{|\sharp(R_0)-\sharp(R)|}
       {\sharp(R)} |\Tr(\rho_0(P_2-P_1))| +
       2\frac{\sharp(R_>)}{\sharp(R)} + \frac{4t_0^2}{M(\delta E)^2}.
       \end{equation}
       
       From the spectrum of $H_0$ we derive
       $$\sharp(R_0)\geq \frac N\pi \left(\frac{\nu_1}{t_1}+\frac{\nu_2}{t_2}\right)\Delta E-1,$$
   and, from the results in section \ref{sec_spectrum}, we have
   $$|\sharp(R_0)-\sharp(R)|\leq 2.$$
   Likewise we can obtain the upper bound
   $$\sharp(R_>)\leq \frac N\pi \left(\frac{\nu_1}{\sqrt{t_1^2-(|E|+\Delta E)^2}}
   +\frac{\nu_2}{\sqrt{t_2^2-(|E|+\Delta E)^2}}\right)\delta E+2. $$

   Using the previous estimates and choosing the cutoff such that
   $$\delta E\to 0, \mbox{ but } N(\delta E)^2 \to\infty, \mbox{ when }N\to\infty,$$
   e. g. $\delta E=N^{-1/3}$, we obtain from (\ref{mainuneq})
   $$  \lim_{N\to\infty}\Tr((\rho-\rho_0)(P_2-P_1))=0,$$
   as stated before.

 \end{document}